# Multi-fidelity prediction of fluid flow and temperature field based on transfer learning using Fourier Neural Operator

Yanfang Lyu[a,b], Xiaoyu Zhao[b], Zhiqiang Gong[b,*], Xiao Kang[c], Wen Yao[b]

[a] State Key Laboratory for Turbulence and Complex Systems & Center for Applied Physics and Technology, College of Engineering, Peking University, Beijing 100871, China

[b] Defense Innovation Institute, Chinese Academy of Military Science, Beijing 100071, China

[c] Institute of Advanced Structure Technology, Beijing Institute of Technology, Beijing 100081, China

[*] Corresponding author.

Email address: gongzhiqiang13@nudt.edu.cn

**Abstract**

Data-driven prediction of fluid flow and temperature distribution in marine and aerospace engineering has received extensive research and demonstrated its potential in real-time prediction recently. However, usually large amounts of high-fidelity data are required to describe and accurately predict the complex physical information, while in reality, only limited high-fidelity data is available due to the high experiment/computational cost. Therefore, this work proposes a novel multi-fidelity learning method based on the Fourier Neural Operator by jointing abundant low-fidelity data and limited high-fidelity data under transfer learning paradigm. First, as a resolution-invariant operator, the Fourier Neural Operator is first and gainfully applied to integrate multi-fidelity data directly, which can utilize the scarce high-fidelity data and abundant low-fidelity data simultaneously. Then, the transfer learning framework is developed for the current task by extracting the rich low-fidelity data knowledge to assist high-fidelity modeling training, to further improve data-driven prediction accuracy. Finally, three typical fluid and temperature prediction problems are chosen to validate the accuracy of the proposed multi-fidelity model. The results demonstrate that our proposed method has high effectiveness when compared with other high-fidelity models, and has the high modeling accuracy of 99% for all the selected physical field problems. Significantly, the proposed multi-fidelity learning method has the potential of a simple structure with high precision, which can provide a reference for the construction of the subsequent model.

## 1. Introduction

Perception of the full state, including flow field and thermal information, is crucial for health monitoring, state analysis, and system design such as aerospace and ocean engineering. Fluid flow and temperature distribution analysis of aerospace and marine engineering devices, have been explored through both field experiments and FEA numerical simulations by lots of researchers [1-6]. As for field experiments, even though the experimental results are generally with high authenticity, many fluid and structural engineering experiments, such as aircraft designation [2] and ocean riser explorations [3], are too expensive to conduct for most of the researchers due to the expensive laboratory equipment and high labor costs. In addition, in some particular situations, such as satellite thermal layout problems, only partial data of measurement points could be obtained due to lots of engineering constraints. While FEA numerical simulations have limited device cost, and can obtain satisfactory results by selecting the appropriate calculation methods including the finite difference method and finite element method [5-6]. However, such a method requires high-precision mesh to generate high-fidelity data, which is time-consuming. To deal with the challenges of the above analysis methods, surrogate-based models were proposed and utilized in recent years [7].

Instead of providing experimental or theoretical results, surrogate modeling technologies try to provide approximating results through machine learning models and have been successfully applied in numerous scientific and engineering studies [8-10]. The construction of a surrogate model can be accomplished using various popular and effective methods including polynomial regression [11-12], polynomial chaos expansions [13], Kriging model [14], support vector machine [15], random forests [16], Artificial Neural Networks [17-18], and et al. The above traditional methods can directly construct surrogate models using available data, nevertheless, have the defects of limited model representation ability and insufficient model accuracy for strong nonlinear systems. Recently, deep models with multiple layers and better representational ability have shown their potential in solving high-dimensional and strong nonlinear problems. Among these deep models, convolutional neural networks (CNNs) [10,19,20] are the representative ones that can extract both the local and global features and present good performance in the reconstruction and prediction of fluid flow and temperature fields [21-22]. While the performance of CNNs is guaranteed by large amounts of training data for avoid the appearance of overfitting, and has poor mesh migration processing ability. Compared with CNN-based models, Fourier neural operator (FNO) proposed by Li et al in 2021 [23] can provide a more stable and better performance owing to its structural regularization effect [24]. Therefore, this paper introduces and improves the FNO, according to the application characteristics, as the surrogate model for engineering project prediction.

As a general problem occurred in deep neural networks, including the FNO architecture, large amounts of high-quality data are required to train a high-accuracy and reliable deep model. However, the generation of high-fidelity data is usually time-consuming, or even unaffordable in some scenarios where physical experiments should be carried out. Factually, the usefulness of the neural network surrogate models depends on the balance between the training costs and the computational benefits generated by deploying the surrogate model. Therefore, it is critical to reduce the data calculation cost or to solve the difficulty of obtaining high-precision data. Taking into account the above matter, the multi-fidelity modeling method, realized by the abundant low-fidelity data and sparse high-fidelity data, has been regarded as a potential and effective measure to reduce the data acquisition costs [12,25,26]. Among them, low-fidelity data requires low acquisition costs but also low precision. While high-fidelity data with high precision are obtained from physical experiments or high-precision numerical simulation, which is expensive and time-consuming to obtain.

Using the multi-fidelity modeling method, the complementary information of low-fidelity and high-fidelity data is utilized to improve the modeling training efficiency and reduce the modeling dependence on high-fidelity data. Traditionally and commonly used multi-fidelity modeling methods include Co-kriging [27-28] based on Gaussian distribution and multi-level Monte Carle [29-30], but have the limitations of dealing with high-dimensional and strong-nonlinear problems. Above two problems can be well tackled by deep neural networks. Recently, deep neural networks have been properly applied to construct multi-fidelity models [12,25,31]. Song et al. build a multi-fidelity model that involves three training steps, i.e., a large set of low-fidelity data and two groups of the same high-fidelity data were successively trained [12]. Chen et al. designed three kinds of relationships between low-fidelity and high-fidelity data, and took full advantage of abundant low-fidelity data [25]. Zhang et al. constructed the complex mapping between one low-fidelity model and two high-fidelity models with paratactic position, and treated the output of the former as the input of the latter [31]. A similar multi-fidelity model was established by Meng et al, but was embedded the effective physical information [32]. Nevertheless, most of the existing multi-fidelity modeling methods introduce an “additional layer” or “bridge function” between the low-fidelity model and the high-fidelity model or generally contain three or more fidelity models, meaning that it is still cumbersome and time-consuming [12]. The reason for adding the “bridge function” is that, for deep models such as CNNs, low-fidelity and high-fidelity data utilized in modeling are generated by coarse mesh and fine mesh, respectively, based on the simulation calculation, resulting in a different number of network weights for constructing two-dimensional data. These relatively complex network frameworks can be simplified by FNO.

Considering the merits of both the FNO and multi-fidelity learning, this paper proposes a novel multi-fidelity learning method based on transfer learning by applying the concept of non-gridded FNO, which can directly transfer the neural network training parameters of the low-fidelity model to the high-fidelity model without any "additional layer" or "conversion joint" of the conversion grid size. This non-gridded FNO migration approach simplifies the modeling structure for coupling multiple-fidelity data, and reduces the dependency on high-fidelity data. This novel multi-fidelity learning method would construct a model that can accurately predict the high-fidelity data, which is called the FNO-based multi-fidelity model (FNO-MFM). To validate the modeling effectiveness of FNO-MFM in global fluid flow and temperature distribution prediction, we apply three engineering projects, i.e., temperature field, airfoil flows, and laminar single-cylinder wake, which contain complex physical information or strong nonlinear dynamics.

To be summarized, this work makes the following contributions.

1. A novel multi-fidelity learning method is proposed based on transfer learning by utilizing the non-grided performance of FNO, which greatly simplifies the neural network structure and solves the dependence of the training process on abundant high-fidelity training samples.
2. An improved Fourier neural operator framework is developed to accommodate the nonlinear flow/temperature fields and to construct the high-precision physical field surrogate model.
3. Three common engineering projects corresponding to three typical nonlinear flow fluid and temperature field prediction problems are applied to analyze the modeling performance of the proposed multi-fidelity learning method.

The purpose of this paper is to establish a multi-fidelity surrogate model based on the non-gridded principle of FNO and the concept of transfer learning. Three different physical field prediction cases are considered for verification, and three generators are created or utilized to obtain the required low-fidelity and high-fidelity data. The effect of the fidelity of low-fidelity data and the number of high-fidelity data on the modeling accuracy of the multi-fidelity model is analyzed in detail. And the modeling performance of the multi-fidelity model and high-fidelity model is explored. The network framework and transfer principle of the multi-fidelity surrogate model are described in detail in Section 2. Three reliable data generators are introduced in Section 3. Section 4 indicates the modeling accuracy and timeliness of established multi-fidelity models. Conclusions can be drawn at the end.

## 2. The framework of multi-fidelity learning method based on the Fourier Neural Operator

The Fourier neural operator is well-known for the capacity for better transferability between different mesh divisions. It compensates for the network dependence of the finite-dimensional operator method by generating a set of grid parameters that can be used for different discretization. Meanwhile, Fast Fourier Transform (FFT) can be conducted to realize the non-linear mapping parameterized. Taking into account the grid independence and powerful approximation ability of FNO, this paper introduces and modifies the FNO to construct a novel multi-fidelity surrogate model. In this section, the implementation principle of the proposed novel multi-fidelity learning method with modified FNO is introduced in detail, consisting of the innovative overall framework, the FNO architecture with various input representations, and the multi-fidelity model training with transfer learning.

### *2.1. Overall Framework*

The overall framework of the multi-fidelity learning method shown in Fig. 1 consists of three subparts,

respectively, data generation with different fidelity, pre-training of the low-fidelity model, and fine-tuning of the high-fidelity model. The parameters of the low-fidelity model are migrated as the initial parameters of the high-fidelity model without any changes to the model structure. The relationship is expressed as follows:

$$y_L = \mathcal{F}^L(x_L) \tag{1}$$

$$y_H = \begin{cases} \mathcal{F}^L(x_H), Epoch = 1 \\ \mathcal{F}^H(x_H), Epoch > 1 \end{cases} \tag{2}$$

where $y_L$ and $y_H$ denote the output of low- and high-fidelity data; $x$ is the corresponding input and $\mathcal{F}$ is the functional relationship between input and output. The application background and methods for data generation are described in detail in Section 3. The principle of pre-training and fine-tuning is derived in Section 2.3.

Specifically, different from traditional methods to construct a high-fidelity model with large amounts of expensive high-fidelity data, this novel method requires training a low-fidelity model first by using a large number of easily obtained low-fidelity samples, which is called the pre-training process. Benefiting from the mesh-invariance of FNO, the network parameters of the low-fidelity model can be directly transferred as the initial parameters of the high-fidelity training, and only a small amount of high-fidelity data is required to train the high-fidelity surrogate model that can accurately predict the high-fidelity flow field data under various working conditions.

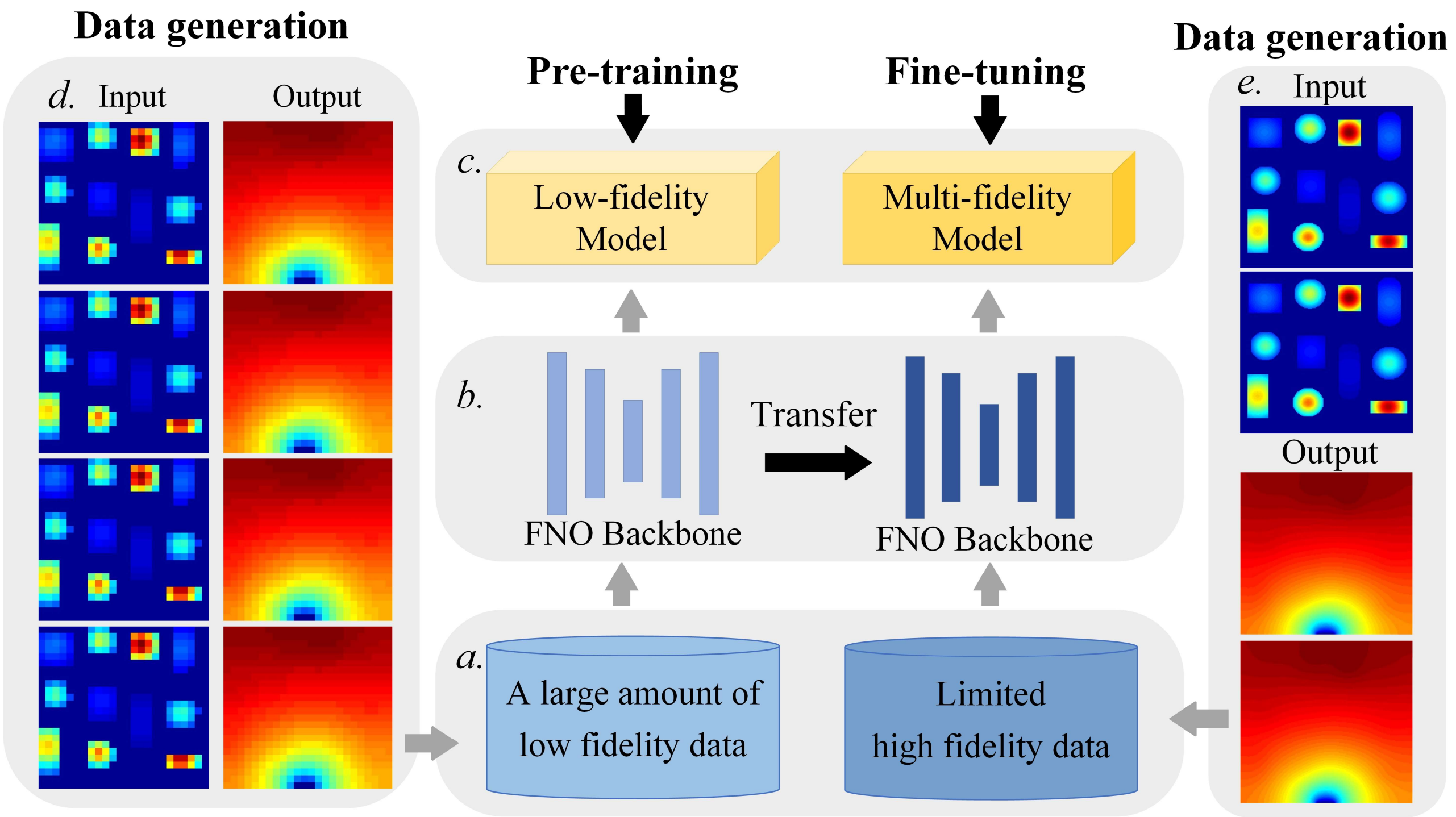

**Fig. 1.** The overall framework of multi-fidelity learning method based on the Fourier Neural Operator.

### *2.2. The Fourier neural operator for field prediction and field reconstruction*

As a primitive algorithm for constructing a multi-fidelity model in this paper, FNO is crucial to the pre-training of the low-fidelity model and the fine-tuning of the high-fidelity model. According to the specific situation of the different physical fields selected, we leverage and improve the original FNO to predict and reconstruct the temperature field and flow fields in Fourier space. As shown in Fig. 2, the improved framework contains an embedding module and multiple Fourier layers. The embedding module is divided into three categories of 2D image data, mask, and MLP-CNN, corresponding to the processing modes for different inputs of three physical fields. The input module should be taken as the initial state of the multiple Fourier layer. Fast

Fourier Transform for each Fourier layer is realized, and then the nonlinear mapping training of network parameters is processed. The superposition of multiple Fourier layers increases the accuracy of physical field reconstruction.

*2.2.1. Embedding for input end*

In Fig. 2 (a), cases I-III represent the embedding modules of temperature field, airfoil flow, and laminar single-cylinder wake, respectively. For FNO modeling, these three engineering examples have the input parameters of a 2D heat source distribution field, 1D combination conditions, and 1D velocity scatter, which are covered in detail in Section 3.

**2D image data.** 2D heat source distribution is regarded as the 2D image data, which can be directly used as the initial input state for the Fourier layer by embedding a convolutional layer with a kernel size of 1×1.

**Mask embedding.** 1D physical quantities including incoming flow velocity and airfoil attack angle can be calculated in combination to generate sub-velocities in $x$- and $y$-direction (i.e., $u_x$ and $u_y$). These two sub-velocities jointing wing shape can be mapped to three two-dimensional mask matrices (i.e., $M_u$, $M_v$, $M_s$) respectively to form the modeling input $a(x)$ with three-way channels. These matrices corresponded to the discrete computational domain (i.e., $D_u$, $D_v$, $D_s$) have the size of ($n_x$, $n_y$). And the coordinate in position ($i$, $j$) is $(D^x_{i,j}, D^y_{i,j})$. Three mask matrices all are divided into two parts including the inside and outside of the airfoil. For $M_s$, the values inside and outside the airfoil are respectively filled with 1 and 0. For $M_u$ or $M_v$, the values inside the airfoil are fixed as 0, which are placed on the positions $\{(x_k, y_k)\}^n_{i=1}$, and the rest positions are filled with $u_x$ or $u_y$, as expressed in follows:

$$M_{i,j} = \begin{cases} \mathbf{0} & \text{if } (D^x_{i,j}, D^y_{i,j})=(x_k, y_k) \text{ for any } k \\ u_x & \text{otherwise, if sub-velocity in x-direction} \\ u_y & \text{otherwise, if sub-velocity in y-direction} \end{cases} \tag{3}$$

**MLP-CNN embedding.** 1D vector **v** contains a series of velocity scatters, as follows:

$$\text{v}=\{v_1, v_2,..., v_m\} \tag{4}$$

where $v_m$ represents the $m$-th velocity scatter. In the proposed network framework, as shown in Fig. 2 (c), the 1D velocity vector is extended by the Multilayer Perceptron (MLP) into the 1D unstructured data. This unstructured data is then reshaped as a 2D low-resolution field which is equivalent to obtaining a coarse grid. Then the series of convolutional layers, instance norm layer, activation function, and interpolation are embedded to generate a high-resolution field $P(x)$, which is used as the initial input of the Fourier layer.

*2.2.2. Fourier layer*

The output of input module $P(x)$ is assigned to the initial input of the multiple Fourier layer $v_1(x)$, as follows:

$$v_1(x) = P(x) \tag{5}$$

Connected with $P$ are four or more Fourier layers (in Fig. 2 (b)) according to the actual application. After completing the training of several Fourier layers, the final output $v_T(x)$ is mapped to the desired output $u(x)$ through the neural network $Q$.

The stacked Fourier layers constitute an iterative architecture $S$.

$$S : v_1 \to v_2 \to ... \to v_T \tag{6}$$

where $v_i$ corresponds to the output of the $i$-th Fourier layer. In each Fourier layer, function iteration, $v_t \mapsto v_{t+1}$, contains a nuclear integral operator $\kappa$, a linear transformation weight $W$ and a nonlinear activation function $\sigma$. The functional relationship between these parameters can be expressed as:

$$v_{t+1}(x) := \sigma(Wv_t(x) + (\kappa(v_t))(x)) \tag{7}$$

$$(\kappa(v_t))(x) = \int k(x,y)v_t(y)dv_t(y) \tag{8}$$

where $\kappa$ and $W$ are parameterized and learnable. The parameterization of the kernel integral operator $\kappa$ can be realized by applying the Fourier transform first and then defining $\kappa$ as the convolution operator in the Fourier space:

$$(\kappa(v_t))(x) = F^{-1}(R \cdot F(v_t))(x) \tag{9}$$

where $\mathcal{F}$ and $\mathcal{F}^{-1}$ represent the Fourier transform and inverse Fourier transform, respectively. $R$ is the parameterization of periodic function $k$ after the Fourier transform.

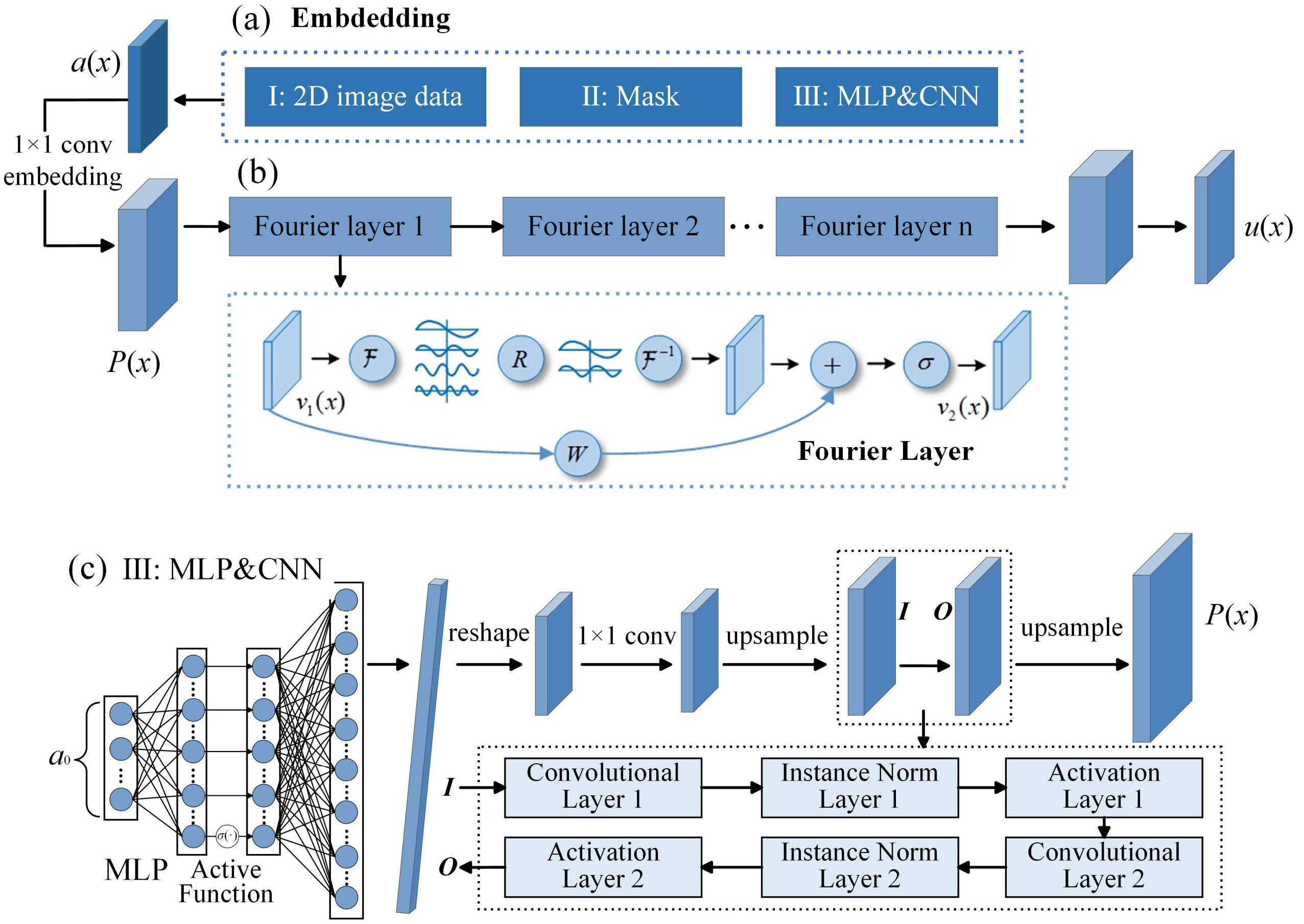

**Fig. 2.** The network framework of Fourier Neural Operator for field prediction and field reconstruction: (a) embedding for input end; (b) Fourier layer; (c) network architecture of example III.

*2.3. Multi-fidelity learning*

The workflow for FNO training on multi-fidelity data is shown in Fig. 3, which involves two phases, i.e., Pre-training and Fine-tuning, as follows:

*2.3.1. Pre-training process*

A large amount of cheap low fidelity data $D^l = \{(P_i^l, U_i^l)\}_{i=1}^N$ under coarse grid are employed to train the low-fidelity model. For low-fidelity network framework of FNO, the predictions $\widehat{U}_i^l$ can be obtained as follows:

$$\widehat{U}_i^l = \mathcal{F}^l(P_i^l), \ \forall i \in \{1,2,\dots,N\} \tag{10}$$

where $\mathcal{F}^l$ denotes the neural operator with learnable parameters. The L1 loss function is employed to optimized the parameters of neural operator in the supervised paradigm, and the loss function $L_{low}$ is described as

$$L_{low} = \frac{1}{N}\sum_{i=1}^{N}\left|\hat{U}_i^l - U_i^l\right| \tag{11}$$

where $U_i^l$ is the corresponding ground truth. To avoid the overfitting, low-fidelity data is divided into two parts that have the ratio of 8:1, one for training and one for validation. Therefore, the optimization network backbone $\mathcal{F}_{opti}^l$ is dependent on the minimum error of validation set. Under the pre-training with low-fidelity data, the neural operator $\mathcal{F}_{opti}^l$ has learn the mapping from $p^l$ to $U^l$, which is not exact but close to the target mapping of high-fidelity data. Since the trained $\mathcal{F}_{opti}^l$ can directly provide the high-resolution predictions with zero high-fidelity samples, it is determined as the initial network to leverage a small amount of high-fidelity data to improve the accuracy of multi-fidelity model. The initial learning rate of pre-training and fine-tuning is chosen as *lr*1 and *lr*2, respectively. Generally, fixing the relationship of *lr*1 > *lr*2 can avoid large changes in model parameters with the higher value of *lr*2, which will destroy features learned from low-fidelity data.

*2.3.2. Fine-tuning process*

A small amount of high-fidelity data $D^h = \{(P_i^h, U_i^h)\}_{i=1}^M$ is used to fine-tune the model, where $N >> M$. Firstly, the pre-trained $\mathcal{F}_{opti}^l$ is transferred as the initial high-fidelity model $\mathcal{F}^h$. Then, calculating the predictive $\widehat{U}_i^h$ as expressed:

$$\widehat{U}_i^h = \mathcal{F}^h(P_i^h), \ \forall i \in \{1,2,\dots,M\} \tag{12}$$

Similarly, $\mathcal{F}^h$ can be trained by minimizing the error $L_{high}$ between predictive value $\widehat{U}_i^h$ and truth value $U_i^h$.

$$L_{high} = \frac{1}{M}\sum_{i=1}^{M}\left|\hat{U}_i^h - U_i^h\right| \tag{13}$$

And the final backbone $\mathcal{F}_{opti}^h$ can be determined by the minimum error of the validation set (the ratio is 1/9) of high-fidelity data. After training of the limited high-fidelity data, we can obtain the multi-fidelity model $\mathcal{M}$, where $\mathcal{M}(*) = \mathcal{F}_{opti}^h(*)$.

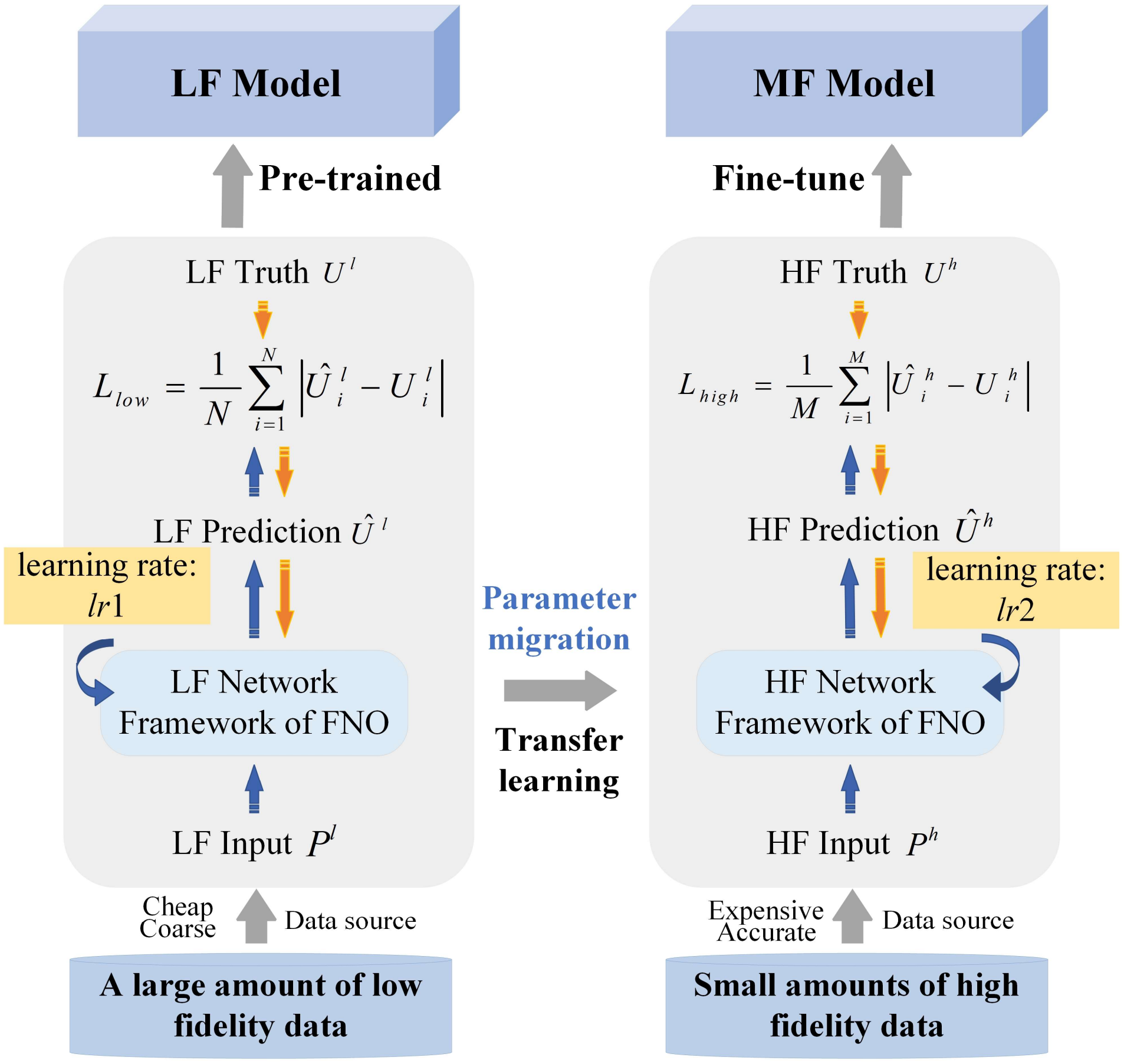

**Fig. 3.** The operating steps of multi-fidelity surrogate model based on FNO.

## 3. Problem statement and data acquisition: temperature field, airfoil flow and single-cylinder wake

We demonstrate the performance of the proposed multi-fidelity learning method for the prediction and reconstruction of temperature and flow fields by using three datasets including the temperature field of heat source layout, 2D airfoil flow, and 2D laminar single-cylinder wake. Three examples, as the typical cases of practical significance for aerospace, aviation, and ocean engineering, are all generated utilizing numerical simulation as follows.

### *3.1. Temperature field prediction of heat source layout*

To increase the functionality and improve the operating speed for industrial products, the development of miniaturization for high-power electronic devices has attracted extensive attention recently. A typical engineering case is the thermal layout of a micro-nano satellite, as shown in Fig. 4 (a)-(b). Nevertheless, the miniaturization with high-power electronic devices will result in heat accumulation for devices, following the service life of the equipment even the serious accidents. In consequence, to guarantee the safety and reliability of the mentioned devices, it is necessary to perform thermal analysis and optimization for the heat source layout. To reduce the cost of expensive experiments or numerical simulations, an temperature field surrogate model based on a deep neural network is advised to construct so as to accurately predict the temperature field of electronic devices [33-34]. As mentioned earlier, high-fidelity data for modeling is not readily available, especially for the satellite thermal layout problem. Therefore, regarding the temperature field as an example to explore the multi-fidelity model has a practical engineering significance.

The purpose of this modeling is to predict the temperature field of the heat source systems at different heat source power. A series of training samples were simulated by fixing the layout position of heat components and changing the heat source power value for each component. In Fig. 4 (d), the two-dimensional square region consists of four adiabatic boundaries except the midpoint of the bottom wall which has a constant heat dissipation hole with the size of $\delta$. The length of four adiabatic boundaries has the value of $L = 0.1$m. And the fixed temperature value at the heat dissipation hole is $T = 298K$. As shown in Fig. 4 (d), eleven heat source components are placed on the square domain. These components are modeled as the shape of circle-like, capsule-like, and rectangle-like with different combinations of length, width, and unevenly distributed temperatures, as shown in Fig. 4 (c) and Table 1. By solving the Poisson equation (i.e., Eq. (14)) with Dirichlet and Neumann boundary conditions, the $2D$ steady temperature field can be simulated in different mesh accuracy.

$$\begin{cases} \frac{\partial}{\partial x}\left(\kappa \frac{\partial T}{\partial x}\right) + \frac{\partial}{\partial y}(\kappa \frac{\partial T}{\partial y}) + \phi(x, y) = 0 \\ Boundary: T = T_0, \kappa \frac{\partial T}{\partial n} = 0 \end{cases} \tag{14}$$

where $T$ and $\kappa$ are the temperature and the thermal conductivity of the layout region, respectively; Denote $\phi_i(x, y)$ as the power intensity of $i$-th heat component with non-uniformly Gaussian distribution [35-36]

$$\phi_i(x, y) = \begin{cases} Q_i \exp(\frac{-\lambda((x-x_0)^2 + (y-y_0)^2)}{r_n^2}), (x, y) \in \Omega_i \\ 0, (x, y) \notin \Omega_i \end{cases} \tag{15}$$

where $\lambda$ is the deviation coefficient and $r_n$ is the radius of the Gaussian heat source. The maximum and minimum value of heat source strength $Q_i$ are $0w/m^2$ and $30000w/m^2$.

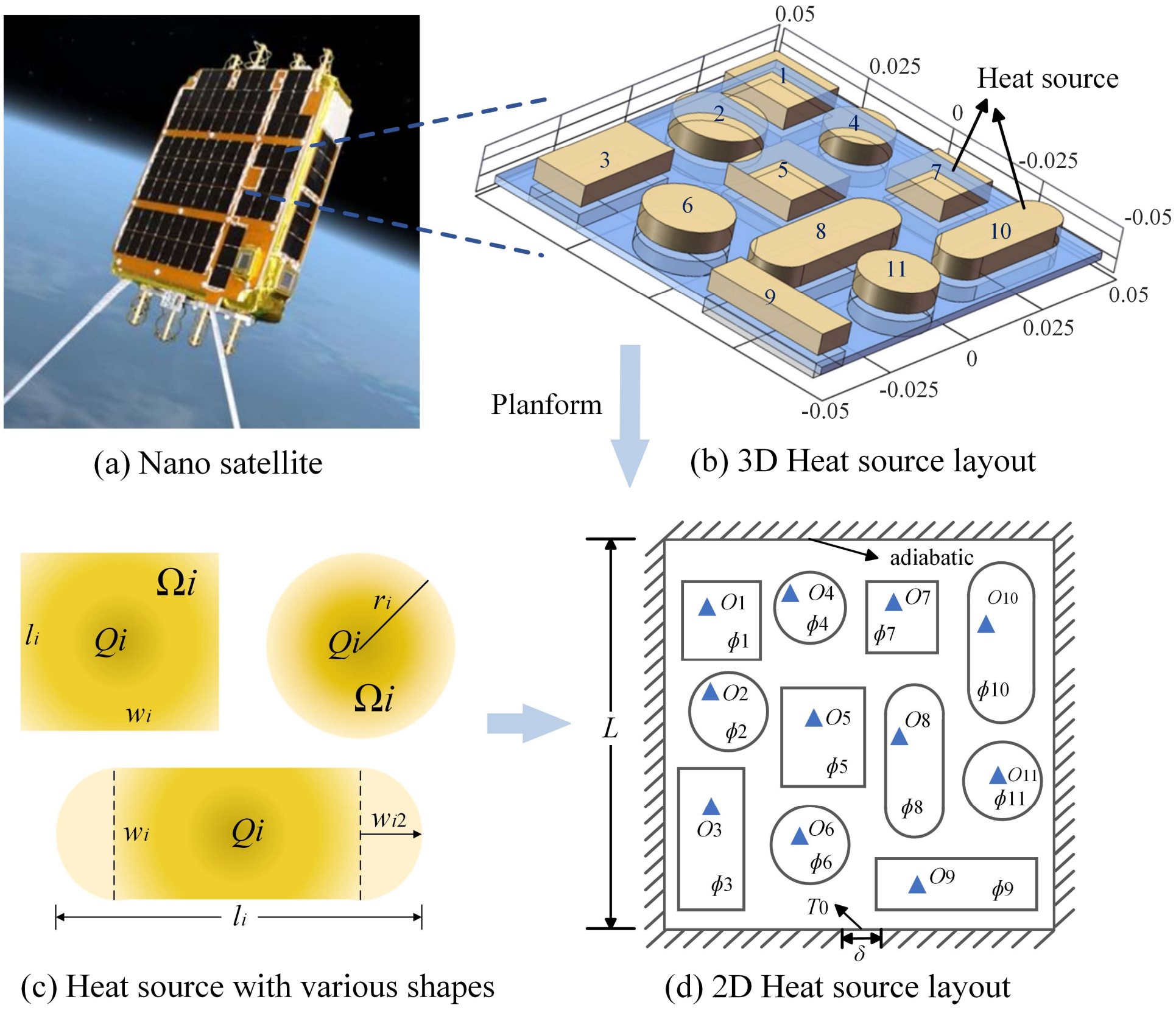

**Fig. 4.** Heat source layout of the nano satellite: (a) nano satellite; (b) 3D heat source layout; (c) schematic diagram of heat source with various shapes; (d) 2D heat source layout and boundary conditions of heat sink.

**Table 1**
The layout characteristics and information of heat source components for Heat Sink.

| No. | Type | Length(m) | Width(m) | Location |
|---|---|---|---|---|
| 1 | rectangle | 0.0182 | 0.0188 | (0.083, 0.085) |
| 2 | circle | 0.0175 | 0.0175 | (0.0535, 0.0845) |
| 3 | rectangle | 0.0273 | 0.0115 | (0.023, 0.089) |
| 4 | circle | 0.0182 | 0.0182 | (0.0855, 0.059) |
| 5 | rectangle | 0.0204 | 0.0163 | (0.05, 0.0585) |
| 6 | circle | 0.0178 | 0.0178 | (0.019, 0.06) |
| 7 | rectangle | 0.0163 | 0.0135 | (0.083, 0.0365) |
| 8 | capsule | 0.0348 | 0.012 | (0.038, 0.0365) |
| 9 | rectangle | 0.0081 | 0.0209 | (0.0165, 0.014) |
| 10 | capsule | 0.0301 | 0.0138 | (0.0805, 0.0135) |
| 11 | circle | 0.0198 | 0.0198 | (0.043, 0.0135) |

The low- and high-fidelity temperature field data corresponding to heat source layout can be, respectively, obtained under the setting of sparse and fine meshes by utilizing the finite difference method (FDM). Specifically, four kinds of temperature fields are divided into 25×25 grids, 50×50 grids, 100×100 grids, and 200 ×200 grids, respectively, as shown in Fig. 5, and generated 2,000 training samples each. Each sample consists of an input and an output, which are respectively the heat source distribution and the corresponding temperature field, as shown in Fig. 5 (*e*)-(*l*). Prescriptively, for the same fidelity, different inputs have the same component layout but dissimilar heat source intensity; meanwhile, data samples for different fidelity are identical except for the resolutions.

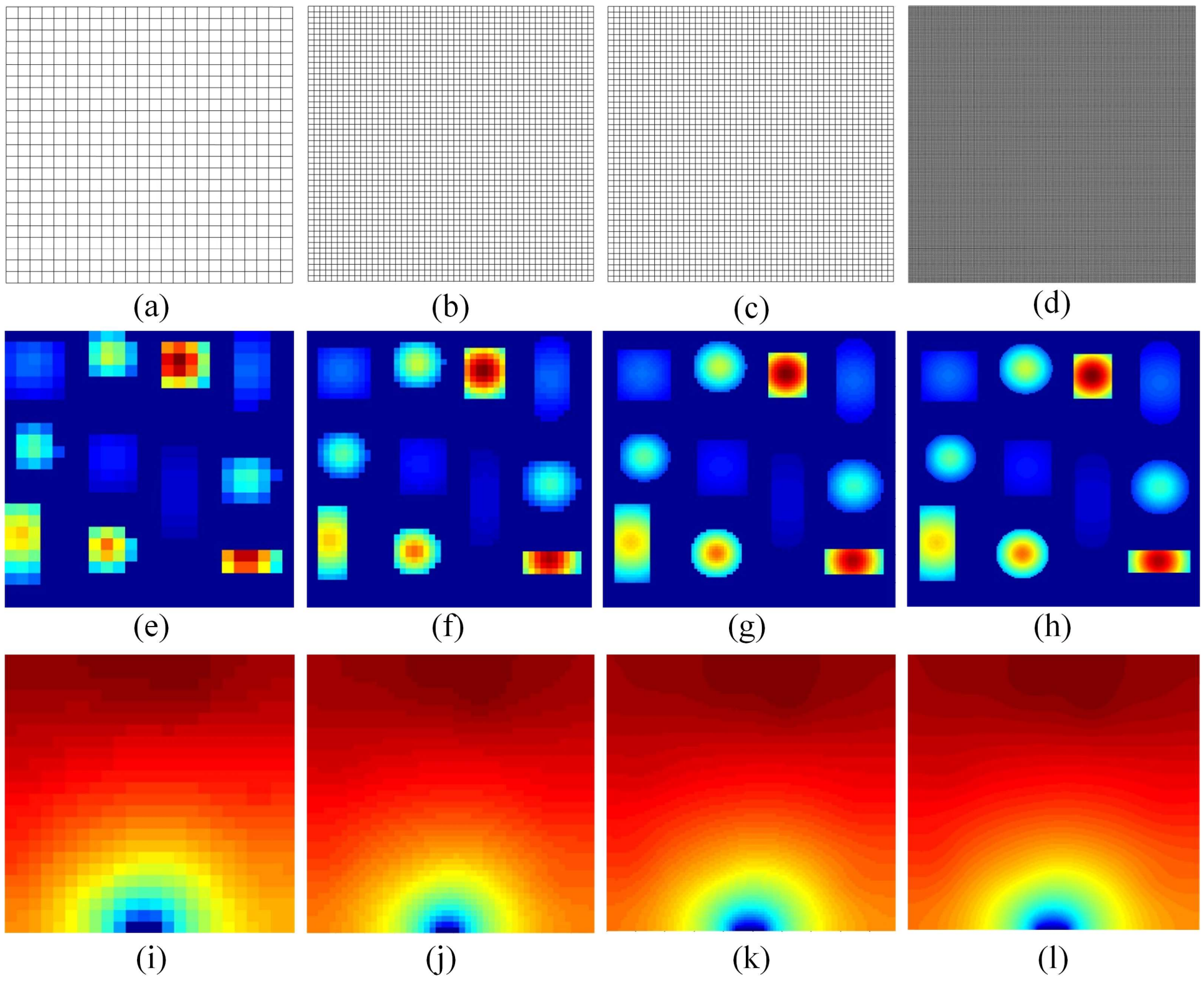

**Fig. 5.** The heat source layout and corresponding temperature field: (a) sparse mesh: 25×25; (b) sparse mesh: 50×50; (c) sparse mesh: 100×100; (d) fine mesh: 200×200; (e) heat source layout: 25×25; (f) heat source layout: 50×50; (g) heat source layout: 100×100; (h) heat source layout: 200×200; (i) temperature field: 25×25; (j)

temperature field: 50×50; (k) temperature field: 100×100; (l) temperature field: 200×200.

*3.2. Airfoil flows*

Under different combinations of flight environmental conditions, an unaltered airfoil will produce a dissimilar airflow pressure field, resulting in various lift/drag applied to the aircraft and unpredictable flight state, as shown in Fig. 6 (a)-(b). The objective of this case is to construct a surrogate model based on the FNO to predict velocity $U$ and pressure $P$ fields around the wing under different incoming airflow velocities $V_\infty$ and attack angle $\alpha$. The constructed model is beneficial to the fast prediction of the airflow fields and has great potential in obtaining optimal flight conditions.

To explore the performance of the proposed novel multi-fidelity model under a specified form of input, three different-fidelity data for airflow fields are generated and trained by setting different grid parameters of the OpenFOAM tool. Following [37], the selected modeling input parameters are $V_\infty$ = 20-40m/s (i.e., $Re$ = 2,000,000-4,000,000) and $\alpha = \pm 10°$, respectively. In addition, the airfoil is fixed as "falcon" so as to simplify the model surrogated. A sufficiently large computational domain of the airfoil data generator would be divided by the unstructured triangular grids. More reasonably, only partial flow fields around the airfoil are interested in and are extracted to train the DNN surrogate. Ultimately, the discrete grid data with a certain fidelity is obtained by interpolation. The interpolation results correspond to three different fidelities that have sizes of 256×256, 128×128, and 64×64. Where the flow field with 256×256 fidelity is selected as the high-fidelity data. Each fidelity generates 2,000 samples, in which one sample contains $S$, $Vx$, and $Vy$ as inputs and corresponding $P$, $U_x$, and $U_y$ as outputs. The subscripts $x$ and $y$ indicate parallel and perpendicular to the airfoil string direction, respectively, as shown in Fig. 6 (c). Fig. 6 (d) shows that $S$ (input channel 1) respects a [0,1] mask for the airfoil shape, 1 being inside and 0 outside. And other two input channels, i.e., $Vx$ and $Vy$, are a [0, $V_\infty x$] and a [0, $V_\infty x$] mask for the shape of the airfoil, respectively, 0 being inside. In addition, three kinds of pressure and velocity fields with different fidelity are shown in Fig. 7.

$$V_{\infty x} = V_\infty \cdot \cos\alpha \tag{16}$$

$$V_{\infty y} = V_\infty \cdot \sin\alpha \tag{17}$$

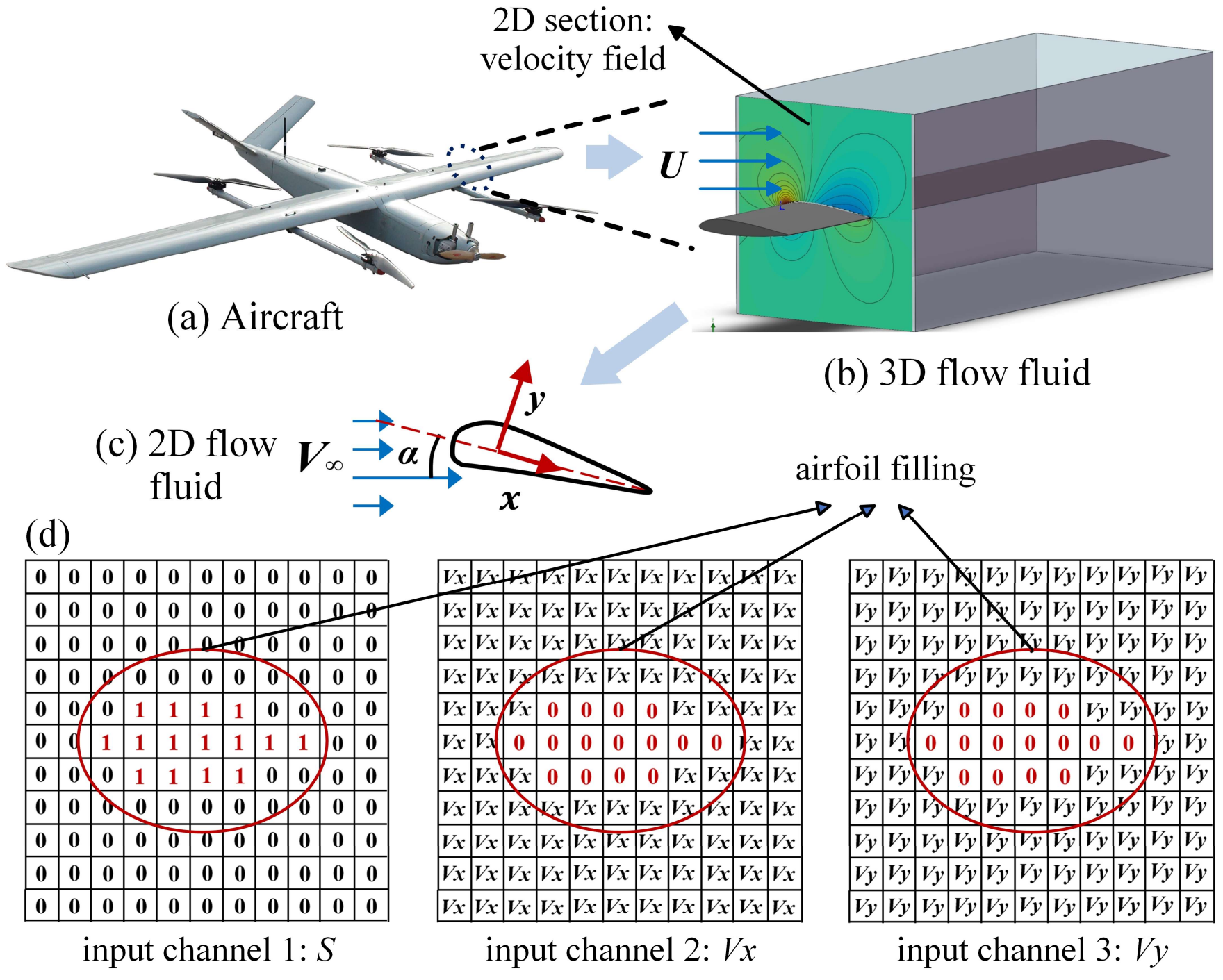

**Fig. 6.** Illustration diagram of incoming flow decomposition and modeling input of 3 channels.

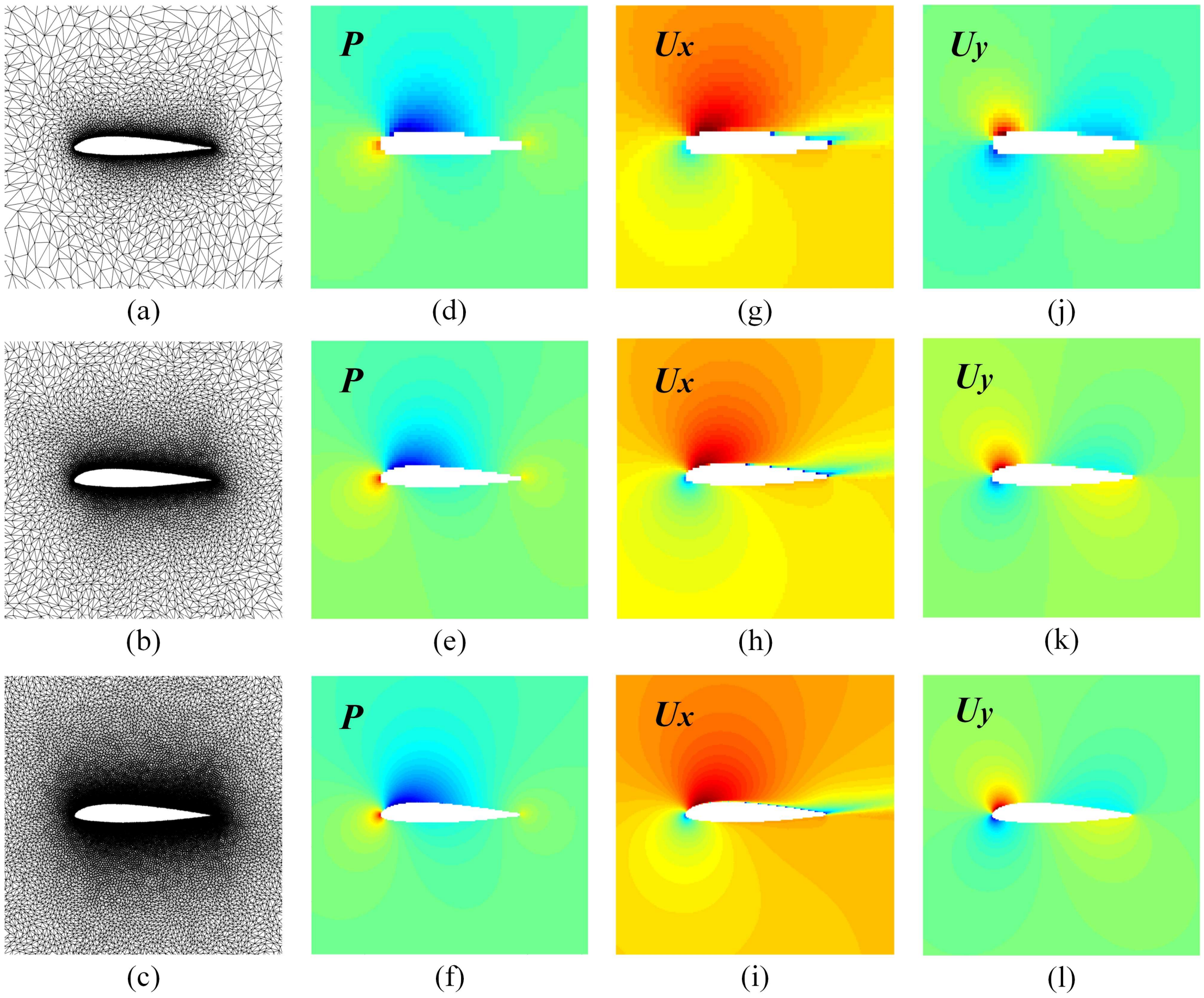

**Fig. 7.** Origin mesh and modeling output of different fidelity: (a) original sparse mesh: 64×64; (b) original sparse mesh: 128×128; (c) original fine mesh: 256×256; (d) $P$: 64×64; (e) $P$: 128×128; (f) $P$: 256×256; (g) $U_x$: 64×64; (h) $U_x$: 128×128; (i) $U_x$: 256×256; (j) $U_y$: 64×64; (k) $U_y$: 128×128; (l) $U_y$: 256×256.

### *3.3. Laminar single-cylinder wake*

Circular cylinders, as a simplified form of engineering structures such as marine risers and piers, are often referred to as basic research objects in engineering theory, as shown in Fig. 8 (a)-(b). In recent decades, a considerable number of scholars were interested in the flow mechanism and characteristics of single or multi-cylinders on account of their typical engineering application, easy generation principle, and abundant flow field information [38-39]. As a typical benchmark in the fluid research field, fluid flow around the cylinder and lift/drag applied to the cylinder have been extensively studied for predictive and reconstructive modeling tasks [40-41]. Based on the above application research background, laminar flow around single-cylinder will certainly be an efficient dataset obtained to explore the accuracy and practicability of the proposed multi-fidelity model. Here, the modeling output is the predictive single-cylinder velocity field, selecting the corresponding evenly spaced scatter velocity as the input parameter of modeling, as shown in Fig. 8.

To ensure the precision degree of the generated high-fidelity data, Lattice Boltzmann Method (LBM) in this paper is utilized for the simulation of a single-cylinder velocity field [42-43] through Matlab 2017b programming. The LBM is a micro discrete model method developed from cellular automata. Different from traditional computational fluid dynamics methods, LBM does not directly solve the macroscopic continuum equation but is based on the microscopic discrete model and makes the overall motion of the microscopic discrete particles conform to the macroscopic motion law [44]. The single relaxation time (SRT) evolution equation is:

$$f_i\left(\vec{x}+\vec{e}_i\delta x, t+\delta t\right)-f_i\left(\vec{x}, t\right)=-\frac{1}{\tau}\left[f_i(\vec{x}, t)-f_i^{eq}(\vec{x}, t)\right] \tag{18}$$

Where $\vec{e_i}$ is discrete speed; $\delta x$ is discrete time step; $\tau$ represents the relaxation time; $f_i(\vec{x}, t)$ and $f_i^{eq}(\vec{x}, t)$ are distribution function and equilibrium distribution function, respectively. The latter can be expressed as:

$$f_i^{eq}(\vec{x}, t)=\rho w_i\left[1+\frac{3\vec{e}_i\cdot\vec{u}}{c^2}+\frac{9(\vec{e}_i\cdot\vec{u})^2}{2c^4}-\frac{3\vec{u}^2}{2c^2}\right] \tag{19}$$

where $\rho$ is macroscopic density; $u$ is macroscopic velocity. $c=\sqrt{3}c_s$, where $c_s$ is sound velocity; $w_i$ represents the weight coefficient. For the discrete velocities of D2Q9 model, $w_i$ is:

$$w_i=\begin{cases}\frac{4}{9}, \vec{e}_0=0, i=0;\\ \frac{1}{9}, \vec{e}_i=\left[\cos(\frac{i-1}{2}\pi), \sin(\frac{i-1}{2}\pi)\right], i=1-4;\\ \frac{1}{36}, \vec{e}_i=\sqrt{2}\left[\cos(\frac{2i-1}{4}\pi), \sin(\frac{2i-1}{4}\pi)\right], i=5-8\end{cases} \tag{20}$$

Necessarily, curve boundary processing is required for single cylinder since the LBM transmits the flow information through virtual particles on the square grids. Allowing for actuarial precision, YMS interpolation format with second order accuracy put forward by Yu et al [45] is referenced in this paper:

$$f_{\bar{i}}^{+}(x_f, t+\delta t)=\frac{1}{1+q}\left[qf_{\bar{i}}^{+}(x_f, t)+qf_i^{+}(x_f, t)+(1-q)f_i^{+}(x_f^{'}, t)\right] \tag{21}$$

To generate high-fidelity data of single cylinder, the cylinder with diameter $D$ = 20 that located at (6$D$+1, 7.5$D$+3) of 30$D$×15$D$ size computational domain is considered, as shown in Fig. 8 (d). The Reynolds number and flow velocity are set to $Re$ = 100 and $U\infty$ = 0.1m/s, respectively. The flow is assumed to be a two-dimensional laminar flow governed by the incompressible Navier-Stokes equation:

$$\begin{cases}\frac{\partial u_i}{\partial t}+\frac{\partial(u_i u_j)}{\partial x_j}=-\frac{\partial p}{\partial x_i}+\frac{1}{\mathrm{Re}}\frac{\partial^2 u_i}{\partial x_j \partial x_j}\\ \frac{\partial u_i}{\partial x_i}=0\end{cases} \tag{22}$$

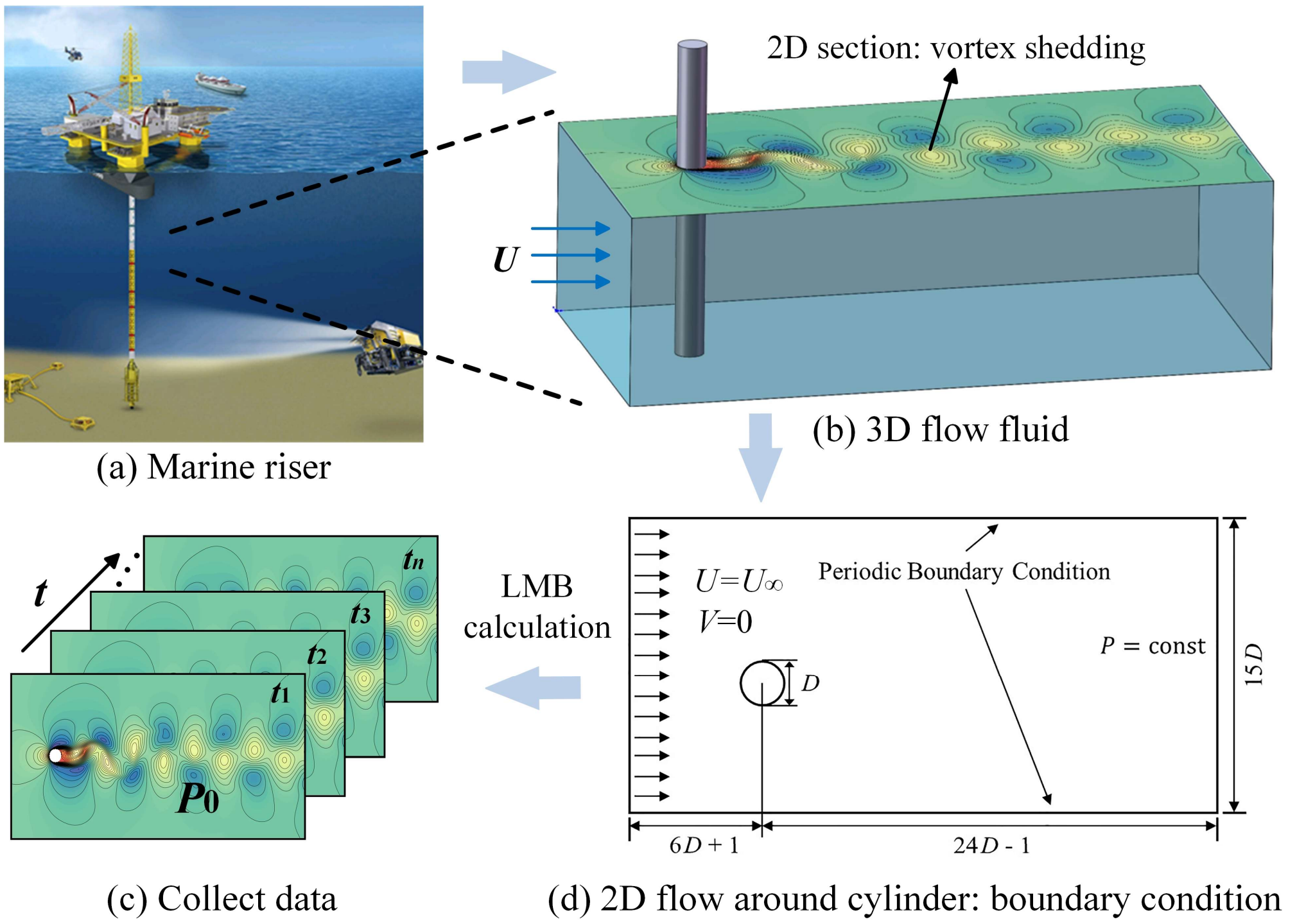

**Fig. 8.** Sketch of single circular cylinder and boundary condition.

Here, the inlet is a horizontal free-stream velocity $U_{\infty}$, and the outlet is set to the constant pressure $P$; periodic boundary conditions are enforced on the upper and lower boundaries; the cylindrical surface adopts the YMS interpolation format as mentioned above. The flow field farther away from the cylinder and vortex street is deleted so that the retained high-fidelity velocity field has the size of 512×256, as shown in Fig. 9 (d). A total of 2,000 flow field snapshots were generated, uniformly obtained from 10 vortex shedding cycles, as shown in Fig. 8 (c). Nevertheless, the velocity field diverges when the computational mesh number is reduced, so the low-fidelity velocity field is selected from the high-fidelity velocity field, as shown in Fig. 9 (a)-(c).

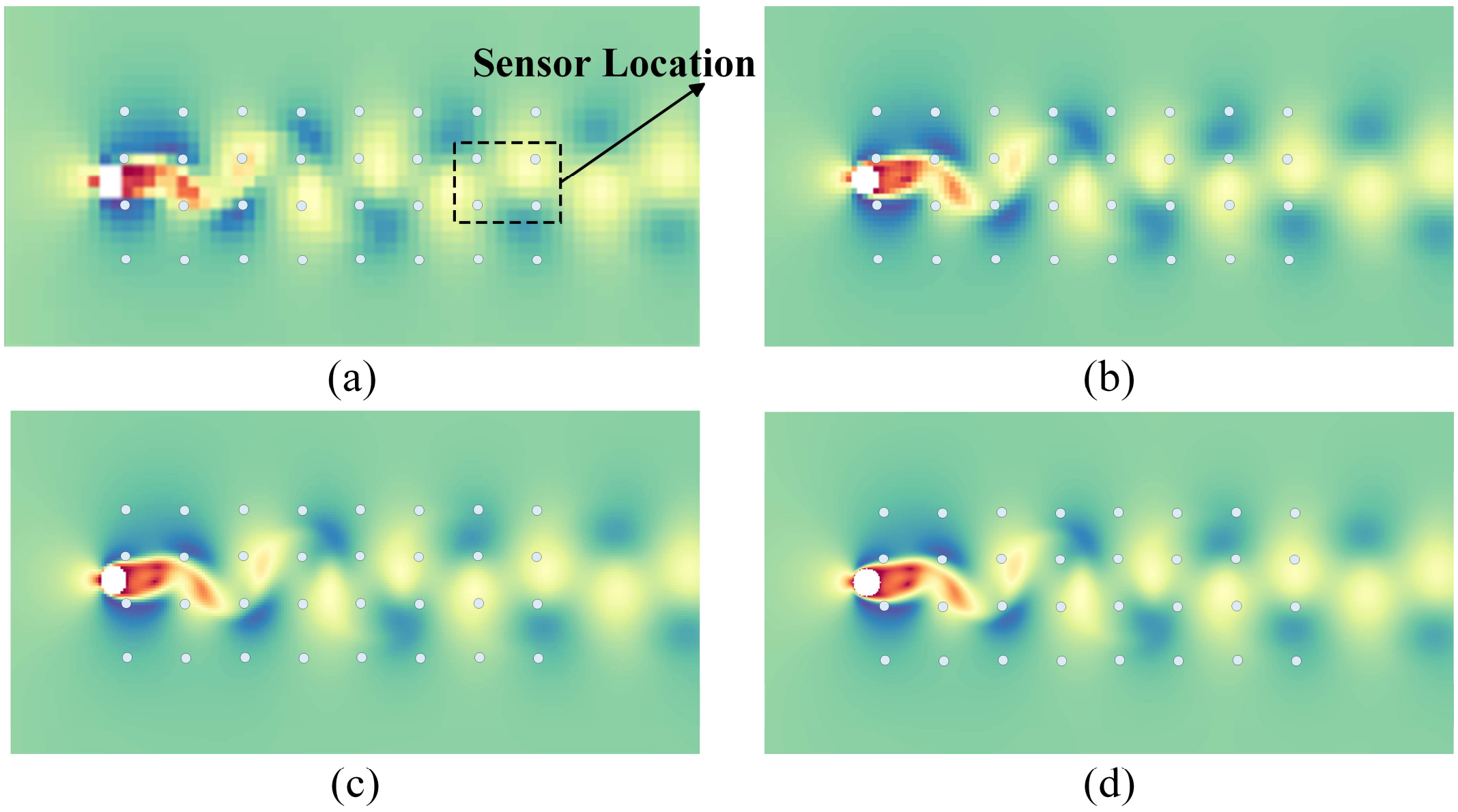

**Fig. 9.** Cylindrical velocity fields with different fidelity: (a) sparse mesh: 64×32; (b) sparse mesh: 128×64; (c) sparse mesh: 256×128; (d) fine mesh: 512×256 (the white circle represent the selected scatter velocity locations).

## 4. Results and discussion

In this section, the proposed novel multi-fidelity learning method (FNO-MFM) is demonstrated for temperature field, airfoil flow, and laminar single-cylinder wake under selected experimental setups and modeling evaluation criteria. As a basic neural network for constructing a multi-fidelity model, the better nonlinear fitting performance of FNO compared with the CNN is validated. The detailed modeling network architecture and pivotal learning rate for FNO-MFM are shown. The effect of the fidelity of low-fidelity data and the amount of high-fidelity data on multi-fidelity modeling is analyzed for three application examples. And the prediction ability and modeling timeless of FNO-MFM are proved in comparison to the high-fidelity model.

### *4.1. Experimental setups and modeling evaluation criteria*

The proposed method is implemented with Pytorch architecture, running on AMD Ryzen 7 PRO 4750U with Radeon Graphics CPU @ 1.70GHz, Nvidia Tesla V100S GPU with 32GB VRAM, and 100 GB RAM.

Mean Absolute Error (MAE) and Mean Relative Error (MRE) are chosen as the metrics to evaluate the performance of various approaches. As a common evaluation metric, MAE is expressed as:

$$\mathrm{MAE} = \frac{1}{N}\sum_{i=1}^{N}\left|\hat{U}_i - U_i\right| \tag{23}$$

where $\hat{U}_i$ and $U_i$ are the predictions and numerical truth results. MRE can avoid the problem of errors canceling each other, so it can accurately reflect the size of the actual prediction error. In contrast with MAE, MRE reflects the magnitude of prediction errors relative to the ground-truth, and is denoted as:

$$\mathrm{MRE}(\%) = \frac{1}{N}\sum_{i=1}^{N}\frac{\left|\hat{U}_i - U_i\right|}{\hat{U}_i}\times 100\% \tag{24}$$

### *4.2. The comparison between FNO and CNN*

As a primitive algorithm for constructing the FNO-MFMs, the strong nonlinear fitting performance of FNO can be verified in comparison with CNNs (traditional CNN and U-net), and the convergence loss of training and validation datasets, for temperature field, airfoil flow, and laminar single-cylinder wake are presented in Fig. 10. These examples have high-fidelity of $S_{HF}$ = 200×200, 256×256, and 256×512, respectively, as recorded in Table 2. The sample number for training, validation, and test, respectively, are 1600, 200, and 200. We train the models for 350 epochs, with a batch size of 16 and an initial learning rate of 0.001.

From Fig. 10, for all the FNO and CNNs modeling, increasing the Epoch will reduce the MAE loss and realize the error convergence. Nevertheless, by comparison, CNNs have the larger loss fluctuation in the early training stage and have higher loss in the final convergence stage. That is, the applied FNO has a faster convergence speed and more accurate prediction ability for all the selected engineering examples. As can be seen in the prediction results in Fig. 11, the temperature and flow fluid fields of FNO show a more accurate forecast and smoother isoline than that of the CNNs. The reason for the better fitting of FNO is that the fast Fourier transform and multilayer Fourier superposition can improve the characterization ability of nonlinear network relations. Overall, the improved FNO has an excellent performance on current field prediction problems.

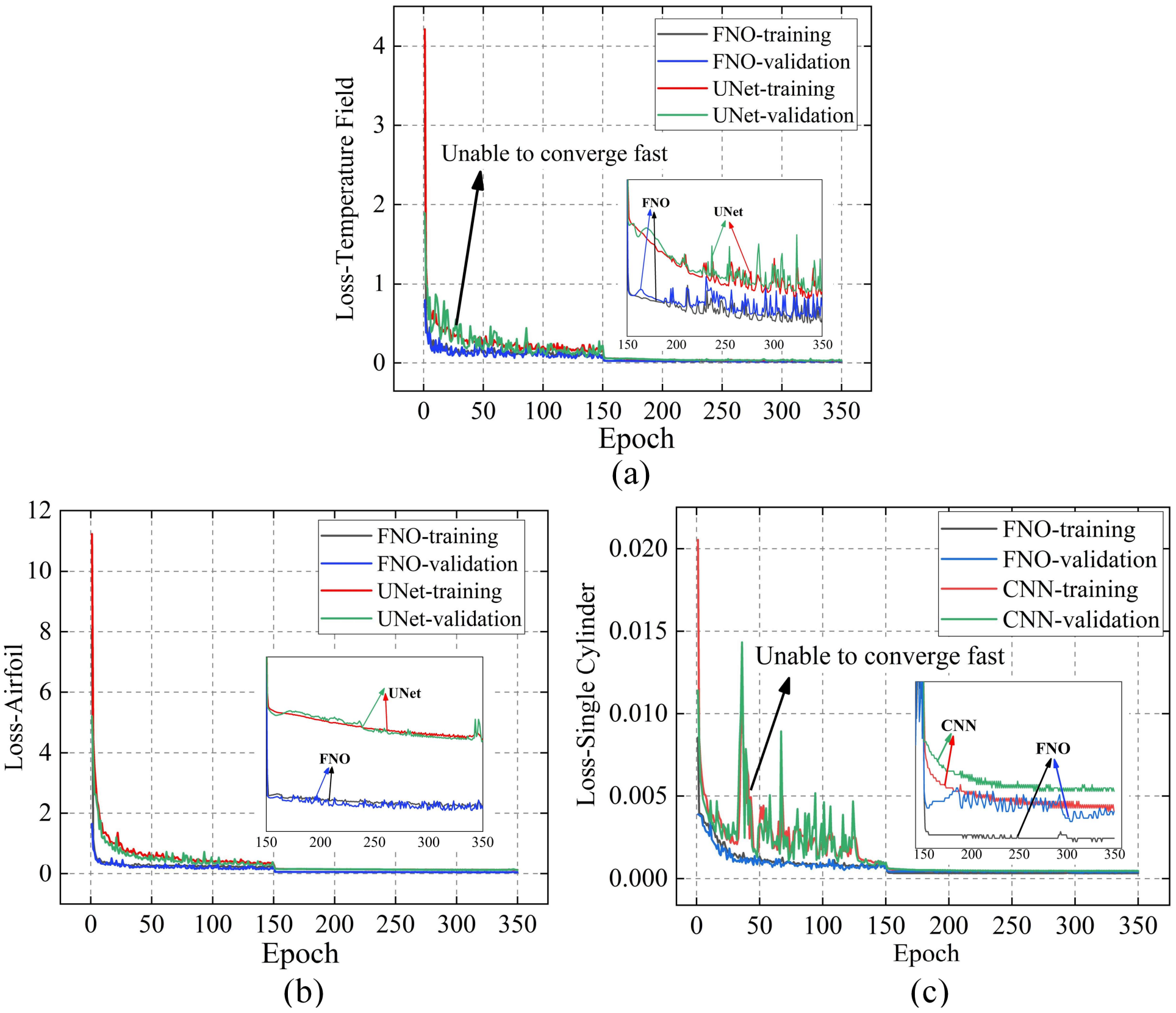

**Fig. 10.** Training and validation loss of FNO, CNN, U-net for three engineering examples: (a) temperature field; (b) airfoil flow; (c) laminar single-cylinder wake.

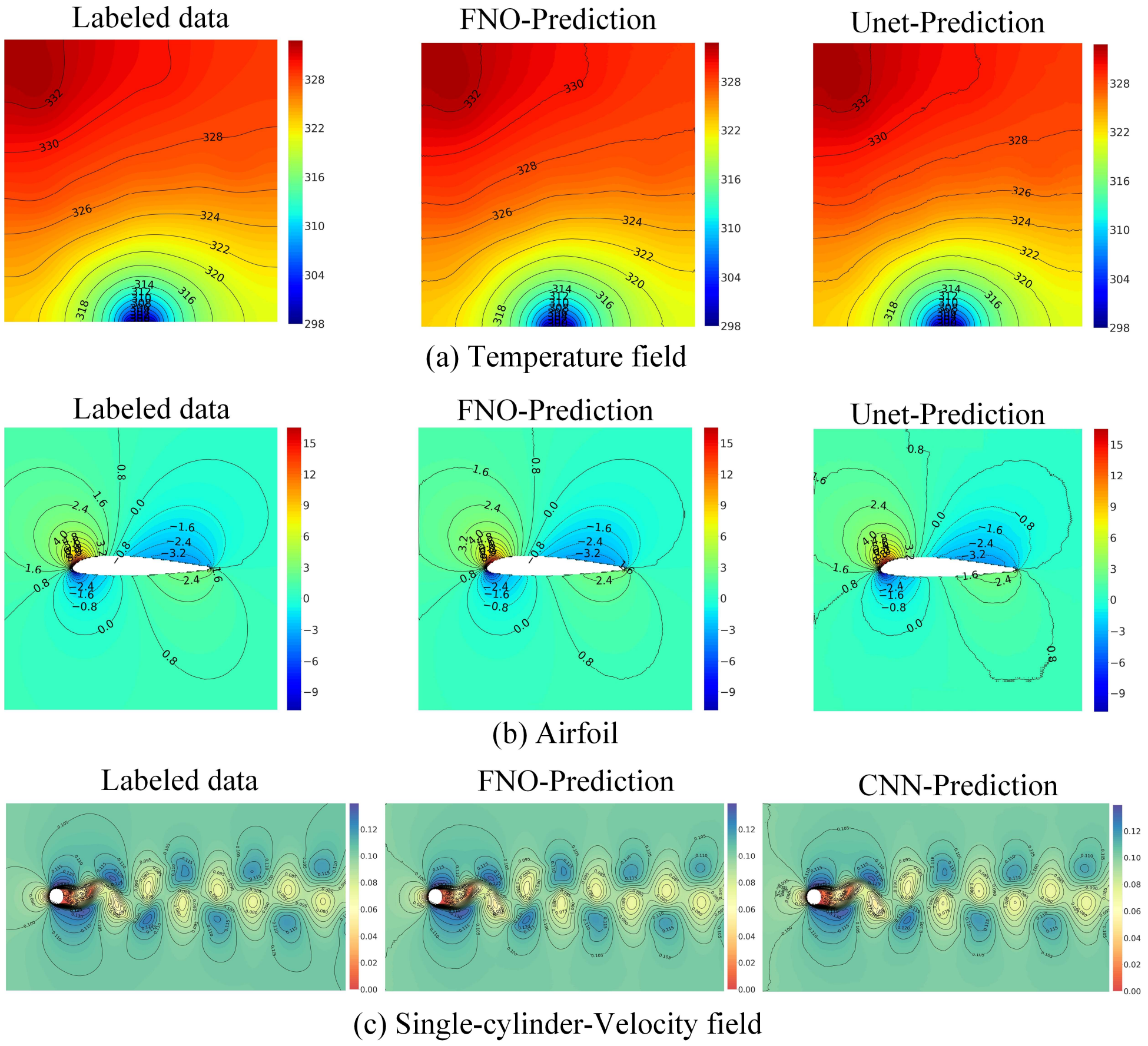

**Fig. 11.** Prediction field of FNO, CNN, U-net for three engineering examples: (a) temperature field; (b) airfoil flow; (c) laminar single-cylinder wake.

### *4.3. Network architecture and learning rate of multi-fidelity model*

As mentioned above, the proposed FNO-based multi-fidelity model can be constructed by a large amount of low-fidelity data and a small amount of high-fidelity data in two stages. For three temperature and flow fluid examples applied in this paper, the training sample number of low- and high-fidelity data, respectively, have the fixed amount of $N_L$=1,600 and a certain range of $N_H$ = {0, 10, 20, 30, 40, 50, 100, 200}. The detailed fidelity and sample number for each considered FNO-MFM are summarized in Table 2.

The network architecture of FNO-MFM frameworks constructed in this paper is described in Table 3. The *GELU* activation function has been used in all the convolution and Fourier Layers. It should be pointed out that, except the "Fully Connected" and "convolution 2/3/5/6" embedded in the single-cylinder example, all the FNO-MFMs adopt the same network framework. In Table 3, $n$ represents the channel number in each layer, where $n_1$=1(example I),3(example II),1(example III); $n_2$=32; $n_3$=128. We train the FNO-MFMs for 500 epochs, with a batch size of 16 using *Adam* optimizer.

The appropriate learning rates set in the two modeling processes are crucial to improve the modeling accuracy of FNO-MFM. In the pre-training process, seven different initial learning rates ($lr1$) are chosen to vary from 1 $\times 10^{-5}$ to 0.1 in equal intervals. The *MultiStepLR* scheduler utilized here establishes the relationship between the learning rate and Epoch. For instance, with 150 and 350 as the cut-off points, the learning rate in three epoch sections is 1, 0.1, and 0.01 times the initial value, respectively. As shown in Fig. 12, with increasing the initial learning rate, the validation loss first decreases and drops up to the minimum at $lr1$ = 0.001, 0.005, and 0.001 for three examples, respectively, and then increases. In consequence, the above three minimums are selected as the initial learning rate of low-fidelity modeling. For high-fidelity training, the corresponding initial learning rate $lr2$ is 0.1 times that of low-fidelity training.

**Table 2**

Fidelity and sample number of each considered FNO-MFM for three engineering examples.

| | Fidelity | | | Sample number |
|---|---|---|---|---|
| | Example I: Temperature | Example II: Airfoil | Example III: Single-cylinder | Training, Validation, Test |
| Low-fidelity $N^{\mathrm{LFC}} \times N^{\mathrm{LFV}}(S_{LF})$ | 25×25 | -- | 32×64 | 1600, 200, 200 |
| | 50×50 | 64×64 | 64×128 | 1600, 200, 200 |
| | 100×100 | 128×128 | 128×256 | 1600, 200, 200 |
| High-fidelity $N^{\mathrm{HFC}} \times N^{\mathrm{HFV}}(S_{HF})$ | 200×200 | 256×256 | 256×512 | 10/20/30/40/50/100 /200, 0.2$N$, 200 |

**Table 3**
Network architecture of FNO-MFMs (corresponding to Fig. 3).

| | Layer | Activation | Input | Output |
|---|---|---|---|---|
| LF model | Learning rate 1 | - | $lr1$ | |
| | Fully Connected | GELU | 28 → 128 → 256 → $N_I^{LFC} \times N_I^{LFV}$ | |
| | Input 1: LF | - | $n_1 \times N_I^{LFC} \times N_I^{LFV}$ | |
| | Convolution 1 (Kernel Size: 1×1) | | | |
| | Convolution 2 (Kernel Size: 3×3) | GELU | $n_1 \times N^{LFC} \times N^{LFV}$ | $n_2 \times N^{LFC} \times N^{LFV}$ |
| | Convolution 3 (Kernel Size: 3×3) | | | |
| | Fourier layer 1/2/3/4 | GELU | $n_2 \times N^{LFC} \times N^{LFV}$ | $n_2 \times N^{LFC} \times N^{LFV}$ |
| | Linear transfer 1 | GELU | $n_2 \times N^{LFC} \times N^{LFV}$ | $n_3 \times N^{LFC} \times N^{LFV}$ |
| | Linear transfer 2 | - | $n_3 \times N^{LFC} \times N^{LFV}$ | $n_1 \times N^{LFC} \times N^{LFV}$ |
| | Output 1: LF | - | $n_1 \times N_O^{LFC} \times N_O^{LFV}$ | |
| Transfer Learning | Migration of parameters | | Fourier parameter, weight, et al | |
| HF model | Learning rate 2 | - | $lr2$ | |
| | Fully Connected | GELU | 28 → 128 → 256 → $N_I^{HFC} \times N_I^{HFV}$ | |
| | Input 2: HF | - | $n_1 \times N_I^{HFC} \times N_I^{HFV}$ | |
| | Convolution 4 (Kernel Size: 1×1) | | | |
| | Convolution 5 (Kernel Size: 3×3) | GELU | $n_1 \times N^{HFC} \times N^{HFV}$ | $n_2 \times N^{HFC} \times N^{HFV}$ |
| | Convolution 6 (Kernel Size: 3×3) | | | |
| | Fourier layer 5/6/7/8 | GELU | $n_2 \times N^{HFC} \times N^{HFV}$ | $n_2 \times N^{HFC} \times N^{HFV}$ |
| | Linear transfer 3 | GELU | $n_2 \times N^{HFC} \times N^{HFV}$ | $n_3 \times N^{HFC} \times N^{HFV}$ |
| | Linear transfer 4 | - | $n_3 \times N^{HFC} \times N^{HFV}$ | $n_1 \times N^{HFC} \times N^{HFV}$ |
| | Output 2: HF | - | $n_1 \times N_O^{HFC} \times N_O^{HFV}$ | |

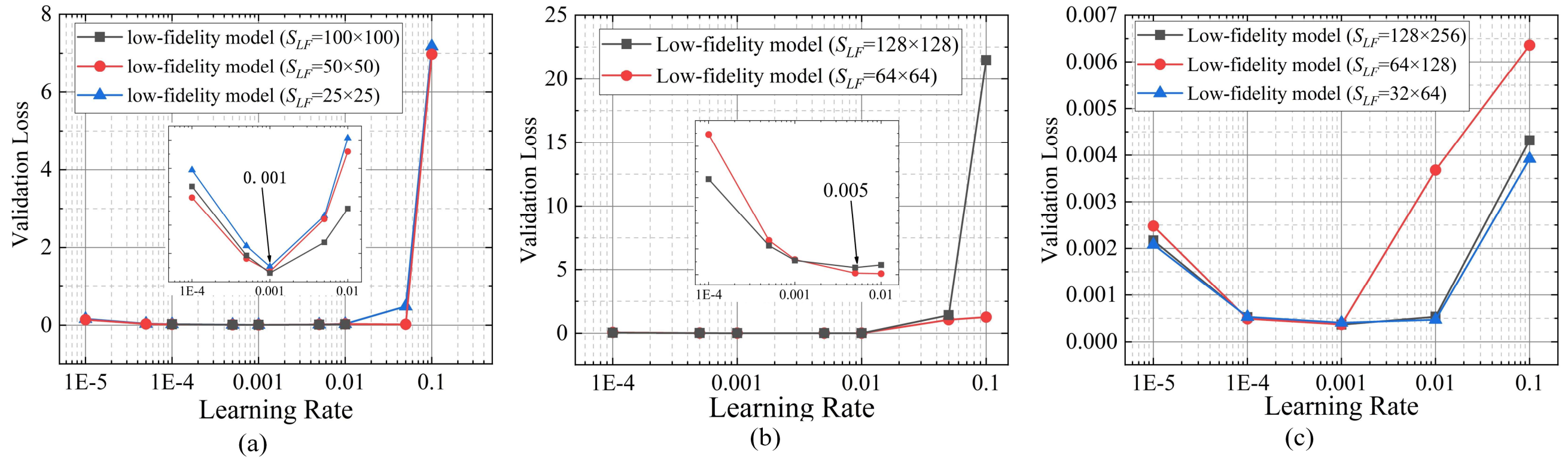

**Fig. 12.** Validation loss versus learning rate: (a) temperature field; (b) airfoil flow; (c) laminar single-cylinder wake.

*4.4. The performance of multi-fidelity learning method based on FNO*

After verifying the superiority of FNO and determining the architecture framework of FNO-MFMs above, we explore the influence of the fidelity of low-fidelity data and the number of high-fidelity data on the prediction performance of FNO-MFM. The fidelity of low-fidelity data directly affects the modeling prediction and generalization of network parameters determined preliminarily by pre-training. Similarly, the more the number of high-fidelity data used in fine-tuning training can provide more reliable physical field information for modeling. To evaluate the modeling accuracy, the MAE and MRE of the test dataset consisting of 200 untrained and unverified are calculated and compared.

#### *4.4.1. Example I: temperature field*

We first consider the temperature field prediction of a nano satellite under different heat source layouts. Fig. 13 illustrates the functional relationship between the modeling test error (MAE or MRE) and the number of high-fidelity data under three different FNO-MFMs and one HFM. The error ratio of FNO-MFM and HFM is shown in Fig. 14. Fig. 15 presents the qualitative comparison of the performance in predicting the temperature field using three FNO-MFMs, with that using HFM. The above visualization figure contains the variables $N_H$ = 0, 10, 50 and 100.

a. The fidelity of low-fidelity data

It can be observed from Fig. 13 that, except modeling with zero high-fidelity data, increasing the fidelity of low-fidelity data from $S_{LF}$ = 25×25 to 100×100 results in a reduction in MAE or MRE, that is, can improve the temperature field prediction of FNO-MFM. More intuitively, the higher fidelity of low-fidelity data has a higher fit between labeled and predicted data and the smoother isotherms in the predicted temperature field, with reference to Fig. 15. However, when the number of high-fidelity samples is zero, the FNO-MFM with $S_{LF}$ = 25×25 has lower MAE and MRE than that of the fidelity of $S_{LF}$ = 50×50. This irrational phenomenon indicates the necessity of using high-fidelity data to fine-tune the model.

b. The number of high-fidelity data

To ensure the consistency of iteration times for each modeling experiment, the existing high-fidelity training data in Table 2 are copied, and increase the number is up to an appropriate value. This incremental operation is reasonable because the true physical information is not increasing. In Fig. 13, with increasing the number of high-fidelity samples from 0 to 200, a prominent drop occurs in both curves of MAE and MRE. This trend is more distinct with fewer high-fidelity samples. Correspondingly, from the visualization of prediction maps, the isotherms are also more reasonably distributed for each FNO-MFM. The increase of high precision samples provides more useful temperature field information for fine-tuning training. From Fig. 15, FNO-MFM with $S_{LF}$ = 100×100 and zero high-fidelity samples can generate a high-resolution temperature field that is close to the real field. The other two FNO-MFMs present relatively weak predictions that heat source shape can be evidently observed in the middle and upper part of the temperature field, however, can still predict the approximate field. The results also indicate that the higher the fidelity of low-fidelity samples, the lower the number of high-fidelity samples required by modeling to satisfy the accuracy requirements. We define this critical minimum as $N_{H\text{-}min}$. For instance, FNO-MFM with $S_{LF}$ = 100×100, 50×50, 25×25, has the approximate $N_{H\text{-}min}$ of 20, 50, and 100, respectively, and has the satisfactory modeling accuracy of 99.97% when $N_H$ = 200.

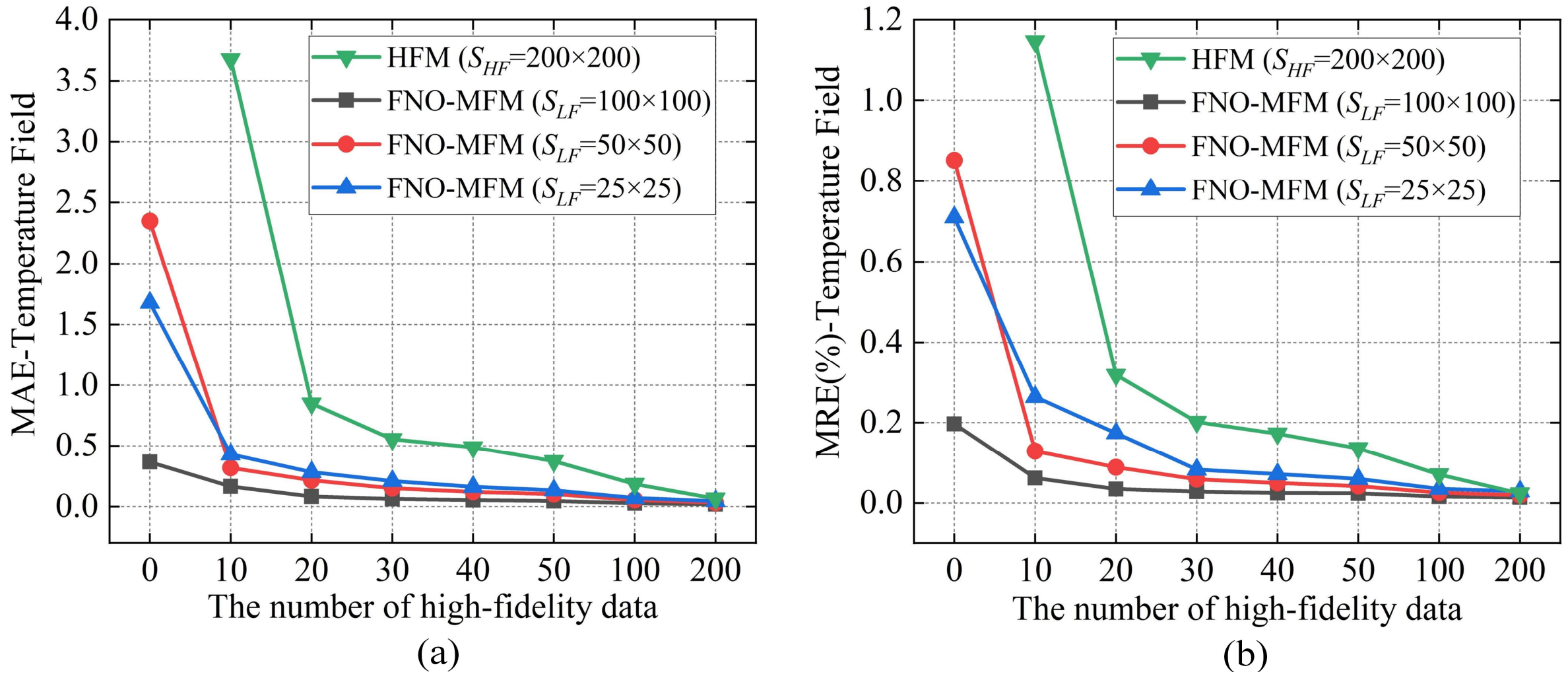

**Fig. 13.** Test error of temperature field versus the number of high-fidelity data: (a) MAE; (b) MRE.

c. Comparison between muti-fidelity model and high-fidelity model

To better reflect the superiority of the proposed FNO-MFM, the MAE and MRE for the high-fidelity model (HFM) with only the high-fidelity samples ($S_{HF}$ = 200×200) is plotted along with corresponding FNO-MFM results for comparison, as shown in Fig. 13. Additionally, the error ratio between the FNO-MFM and HFM based on the test MAE or MRE in Fig. 13 is calculated, as drawn in Fig. 14. In Fig. 13 and Fig. 14, as the number of high-fidelity samples increases from 10 to 100, all the MAE and MRE of FNO-MFMs remain obviously lower than that of the HFM, indicating that the FNO-MFM has better performance than the corresponding HFM. Quantificationally, except for the FNO-MFM with $S_{LF}$ = 25×25 when $N$ = 200 (in Fig. 14 (b)), other error ratios are all lower than 1.0. The minimum value of the MAE ratio and MRE ratio are 0.045 and 0.05, respectively, both of them belong to the FNO-MFM with $S_{LF}$ = 100×100. It means that a large amount of low-fidelity data would carry abundant useful temperature field information although it has limited data accuracy. This is equivalent to the construction of a relatively reliable prediction model with low precision. In addition, with increasing the number of high-fidelity samples, the overall change of error ratio shows an upward trend. This phenomenon is consistent with the prior knowledge that the training results of low-fidelity data may interfere with the accuracy of HFM to some extent. Besides, the other advantage is that FNO-MFM can obtain the high-fidelity temperature field with a certain precision under zero high-fidelity samples owing to its migration function of model parameters. As shown in Fig. 15, the accuracy of FNO-MFM with $S_{LF}$ = 100×100 when $N_H$ = 0 is higher than that of the HFM when $N_H$ = 50. By comparison, FNO-MFM with selected three fidelity all can obtain the precise or approximate distribution of temperature field, however, HFM cannot realize. It should be pointed out that the network parameter of HFM with $N_H$ = 0 is replaced by that of training data with $N_L$ = 1 and $S_{LF}$ = 100×100.

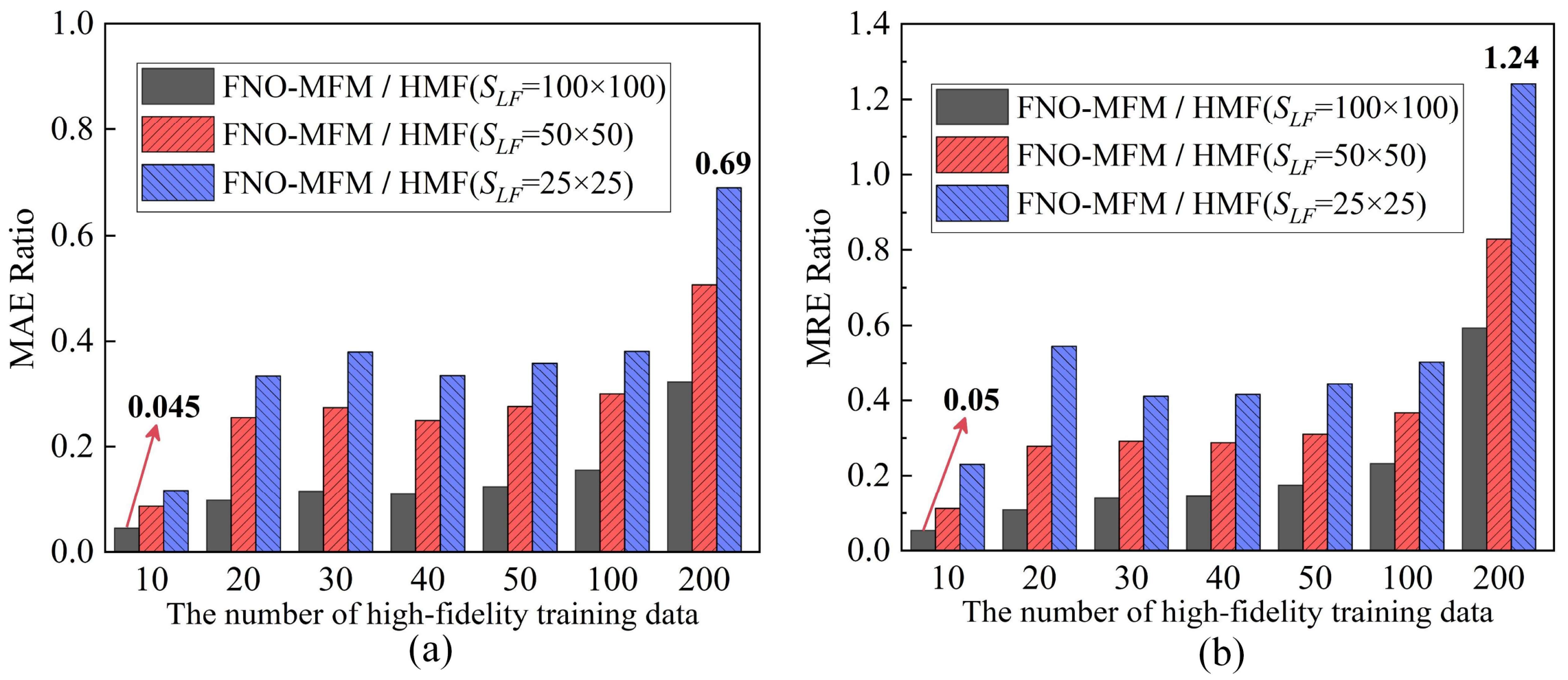

**Fig. 14.** Temperature field--ratio of test error between FNO-MFMs and HFM: (a) MAE; (b) MRE.

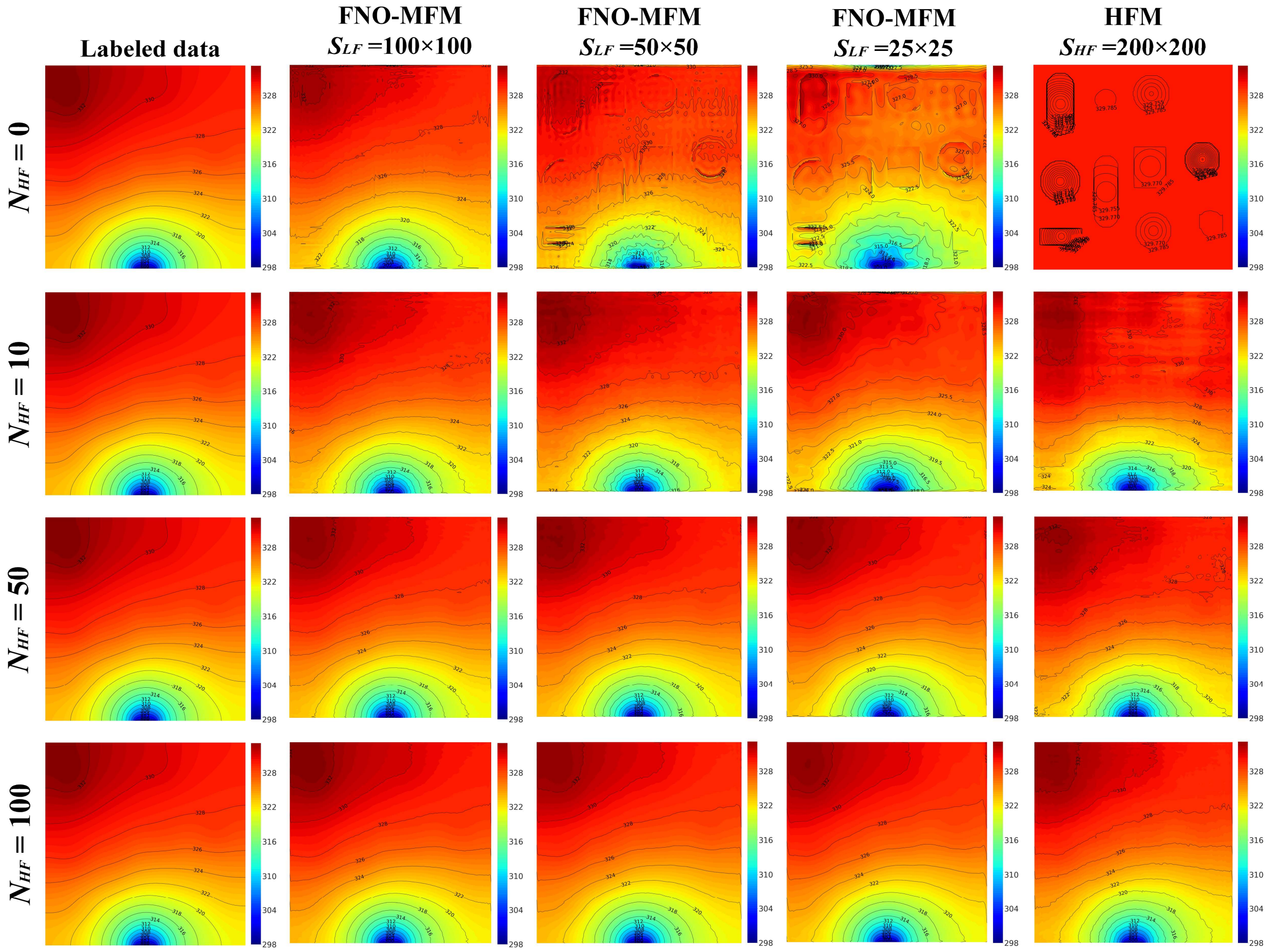

**Fig. 15.** Qualitative comparison of the performance in predicting the temperature field using three FNO-MFMs (FNO-MFM I: $S_{LF}$=100×100 and $S_{HF}$=200×200; FNO-MFM II: $S_{LF}$=50×50 and $S_{HF}$=200×200; FNO-MFM III: $S_{LF}$=25×25 and $S_{HF}$=200×200) and one HFM ($S_{HF}$=200×200).

### *4.4.2. Example II: two-dimensional airfoil wake*

In this section, the prediction of airflow fields including pressure field and velocity field is considered under different combinations of incoming flow velocity and attack in turbulent conditions. Turbulent airflow contains rich field phenomenon and has more challenging for multi-fidelity research.

a. The fidelity of low-fidelity data and the number of high-fidelity data

The MAE and MRE for two FNO-MFMs versus the number of high-fidelity data are plotted in Fig. 16. It should be reminded that the prediction error here is divided into two parts, i.e., pressure and velocity. In Fig. 16 (a)-(b), as $N_H$ increases from 10 to 50, the FNO-MFM with $S_{LF}$ = 128×128 has higher forecasting precision for the pressure field than that of the FNO-MFM with $S_{LF}$ = 64×64. In addition, the MAE and MRE both first drop up obviously and then decrease steadily as $N_H$ increases for the pressure field. Fig. 18-Fig. 20 presents the qualitative comparison between two FNO-MFMs. It can be observed that with increasing the number of the high-fidelity training sample, the difference between the prediction effect of two FNO-MFMs become smaller. For velocity prediction, the variation trend of errors and the comparison of the airflow field follow closely the situation of the pressure field. Different from pressure, the velocity field is divided into two directions in cartesian coordinates as shown in Fig. 6 (c), and corresponding visualization results are displayed in Fig. 18 and Fig. 19. In visualization figures, with the increases in the fidelity of low-fidelity data or the number of high-fidelity data, either the isobar or the isokinetic lines become smooth. Additionally, owing to the more physical information that the pressure field contained, the predicted MRE of the pressure field has high values than that of the velocity results in Fig. 16 (b) and Fig. 16 (d). Overall, the MRE of pressure and velocity field can both reach up to 99% when $N_H$ = 50.

b. Comparison between muti-fidelity model and high-fidelity model

In Fig. 16, within the scope of $N_H$ = 10-50, two FNO-MFMs both have lower MAE than that of the HFM. A quantitative demonstration is shown in Fig. 17, all the MAE ratios have a value of less than 1.0. For the FNO-MFM with $S_{LF}$ = 64×64 when $N_H$ = 10, the MAE and MRE are close to that of HFM under the same amount of high-fidelity data, indicating that data with relatively low-fidelity contains relatively less useful pressure field information, which is more evident in Fig. 17 (a)-(b). For MRE from Fig. 16 (b) and Fig. 16 (d), as $N_H$ = 40 and 50, FNO-MFMs have worse performance compared to the HFM. This is further reflected in Fig. 17 (b) and Fig. 17 (d). This observation is due to the modeling ease of FNO for airfoil field construction.

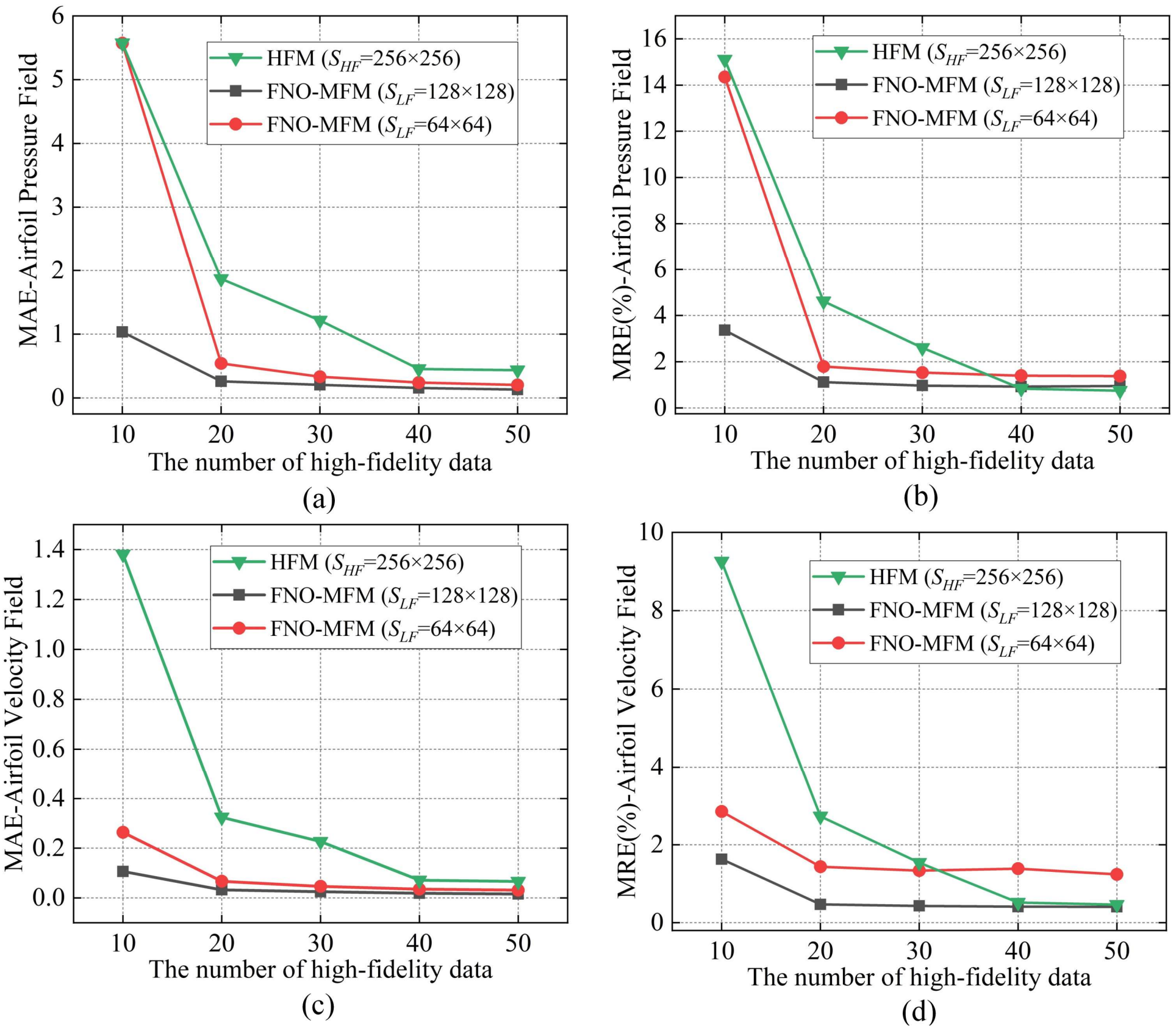

**Fig. 16.** Test error of airfoil flow versus the number of high-fidelity data: (a) MAE-*P*; (b) MRE-*P*; (c) MAE-*U*; (d) MRE-*U*.

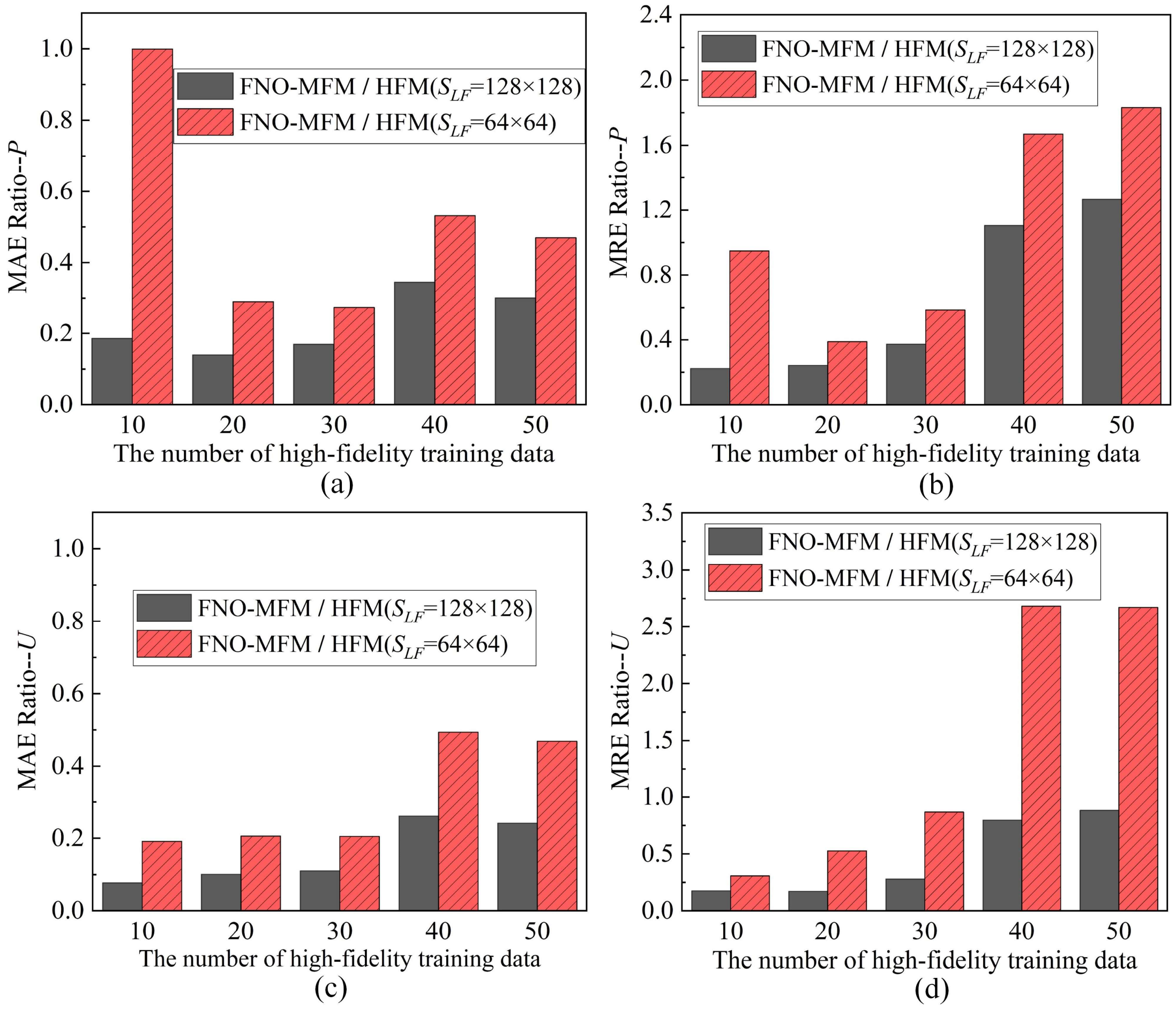

**Fig. 17.** Airfoil wake--ratio of test error between FNO-MFMs and HFM: (a) MAE-*P*; (b) MRE-*P*; (c) MAE-*U*; (d) MRE-*U*.

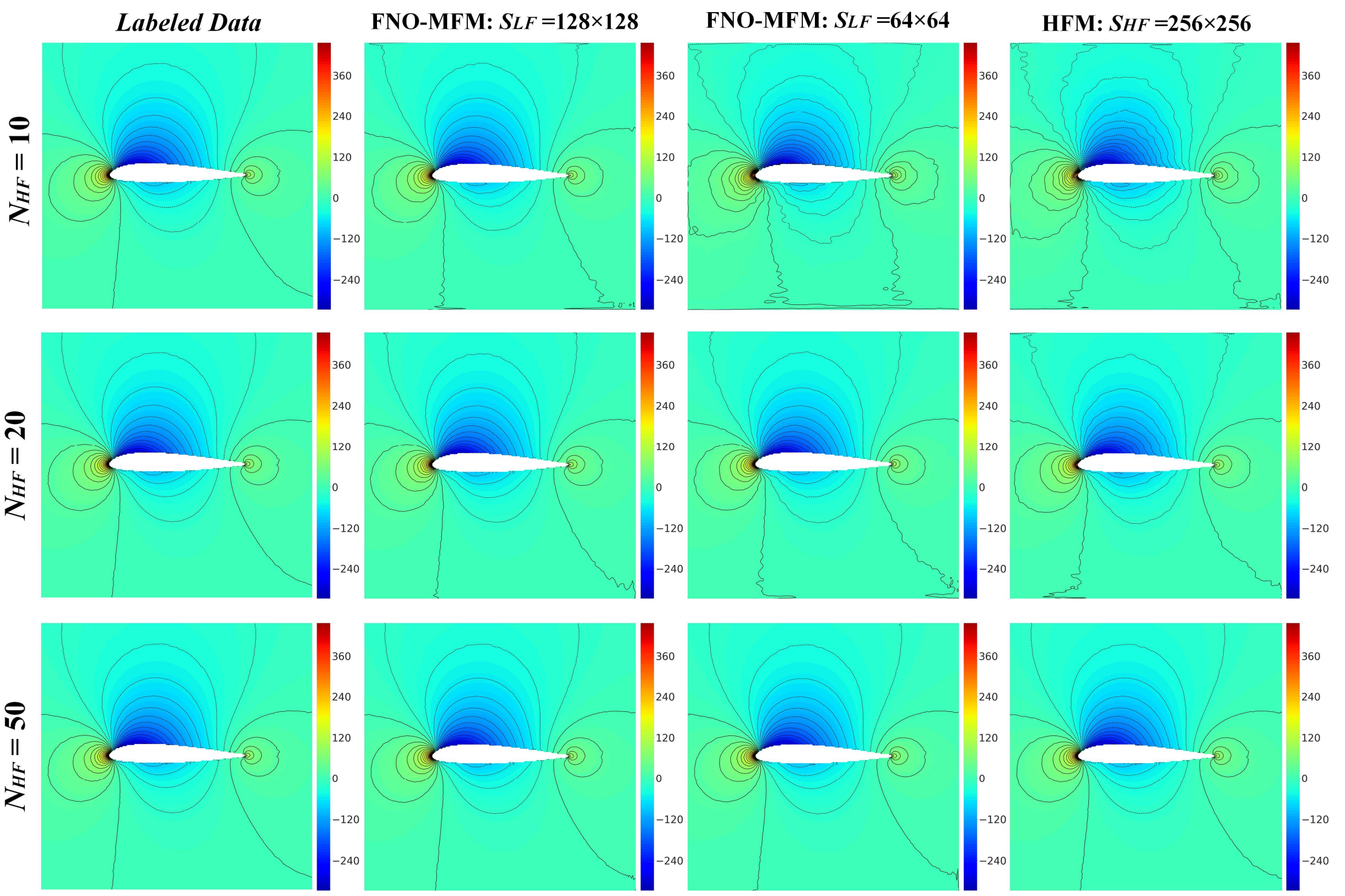

**Fig. 18.** Qualitative comparison of the performance in predicting the pressure field ($P$) of airfoil flow using two FNO-MFMs (FNO-MFM I: $S_{LF}$=128×128 and $S_{HF}$=256×256； FNO-MFM II: $S_{LF}$=64×64 and $S_{HF}$=256×256) and one HFM ($S_{HF}$=256×256).

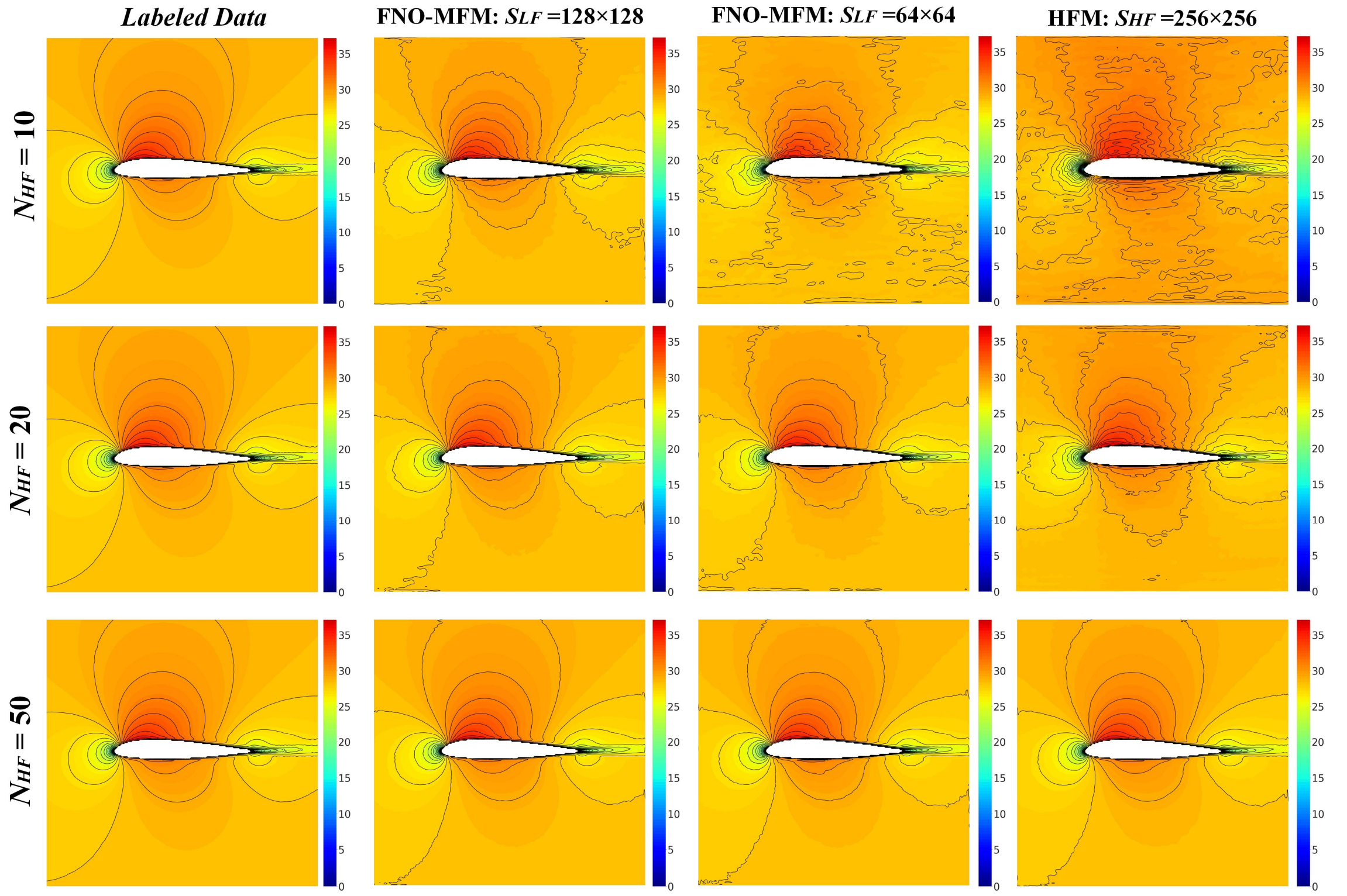

**Fig. 19.** Qualitative comparison of the performance in predicting the velocity field ($Ux$) of airfoil flow using two FNO-MFMs (FNO-MFM I: $S_{LF}$=128×128 and $S_{HF}$=256×256； FNO-MFM II: $S_{LF}$=64×64 and $S_{HF}$=256×256) and one HFM ($S_{HF}$=256×256).

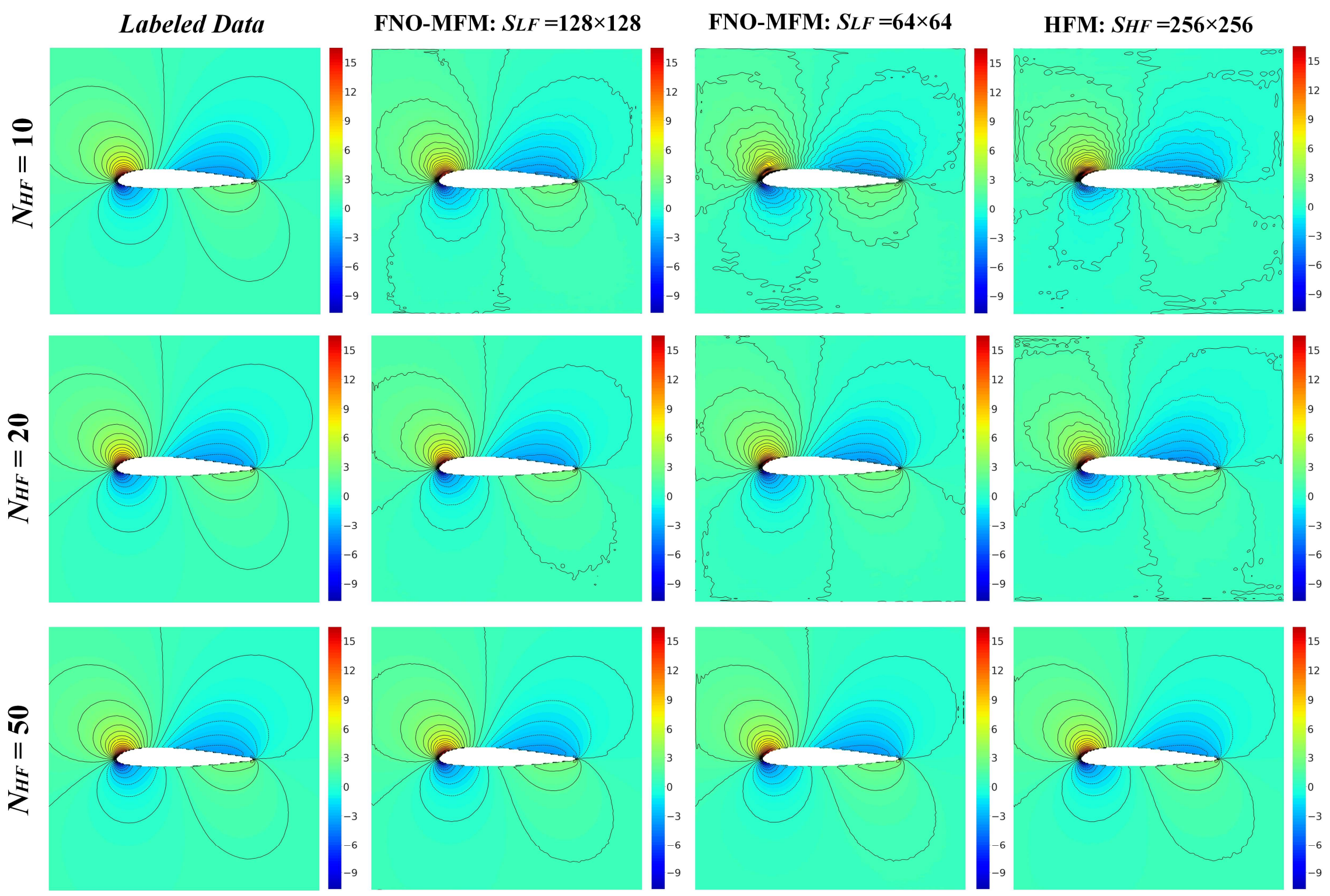

**Fig. 20.** Qualitative comparison of the performance in predicting the velocity field ($Uy$) of airfoil flow using two FNO-MFMs (FNO-MFM I: $S_{LF}$ = 128×128 and $S_{HF}$ = 256×256; FNO-MFM II: $S_{LF}$ = 64×64 and $S_{HF}$ = 256×256) and one HFM ($S_{HF}$ = 256×256).

#### *4.4.3. Example III: laminar single-cylinder wake*

a. The fidelity of low-fidelity data

Similar to the above two examples, Fig. 21 plots the variation trend of MAE and MRE for a single-cylinder velocity field with increasing the number of high-fidelity samples at $Re$ = 100. As expected, as the fidelity of low-fidelity data increases, the prediction error of FNO-MFM shows an obvious drop at the selected number of high-fidelity data except for $N_H$ = 200. For instance, FNO-MFM with $S_{LF}$ = 128×256 and $S_{LF}$ = 32×64, respectively, has the lowest and highest test errors. The former can achieve considerable accuracy as equipped with zero or tiny minority high-fidelity samples. The latter has relatively low effectiveness than that of the other two FNO-MFMs. These phenomena are more evident in Fig. 22-Fig. 24. As shown in Fig. 22, when $N_H$ = 10-50, the error ratios for FNO-MFM with $S_{LF}$ = 32×64 are invariably greater than 0.8, however, the other two FNO-MFMs have error ratios lower than 0.3. In Fig. 23, the cylindrical wake of FNO-MFMs with $S_{LF}$ = 128×256 and $S_{LF}$ = 64×128 both show a shape that closely matches the real field while chaotic wake is observed for FNO-MFM with $S_{LF}$ = 32×64 when $N_H$ = 0-100. Furthermore, smooth, zigzag and chaotic isokinetic lines appear respectively in the above three FNO-MFMs. This lowest accuracy for $S_{LF}$ = 32×64 case is mostly attributed to the large information gap of simulation results between the coarse and accurate single-cylinder velocity field calculated by the LBM.

To quantitatively compare the impact of reduced fidelity on the modeling accuracy for FNO-MFMs, "Error Reduction Rate ($ERR$)" is defined for evaluation, as expressed in Eq. (25).

$$ERR_{H-L}^{MAE} = \left| MH - ML \right| / MH \tag{25}$$

where $ML$ and $MH$ represent the predicted error of FNO-MFMs with lower fidelity and higher fidelity,

respectively. As recorded in Table 4, $ERR_{256\text{-}128}$ has a lower value than that $ERR_{128\text{-}64}$, meaning that the reduction in fidelity is not proportional to the increase in error. It proved once again that low-fidelity data contain less real physical knowledge and even induce interference compared to the real field.

b. The number of high-fidelity data

As the number of high-fidelity data increases, the prediction accuracy of each FNO-MFM follows higher (in Fig. 21). However, the modeling error of adjacent high-fidelity numbers for FNO-MFM with $S_{LF}$ = 128×256 and $S_{LF}$ = 64×128 reduces within 1.35% and 10.3%, respectively. This inconspicuous decrease is due to the high accuracy achieved in the few high-fidelity samples. Corresponded critical minimum $N_{H\text{-}min}$ has the appropriate value of 10 and 50, respectively. Fig. 23 and Fig. 24 illustrate that, although FNO-MFM with $S_{LF}$ = 32×64 can forecast the approximate flow field distribution, the overall prediction effect is not satisfactory. The high fitting precision with the accuracy of 99% in the wake field of FNO-MFMs and HFM when $N_H$ = 200 is associated with the vortex shedding period ($T$ = 200) of a single cylinder, which indicates that training the velocity fields in a certain cycle is sufficient to construct a high precision model. Even too low fidelity will interfere with modeling accuracy.

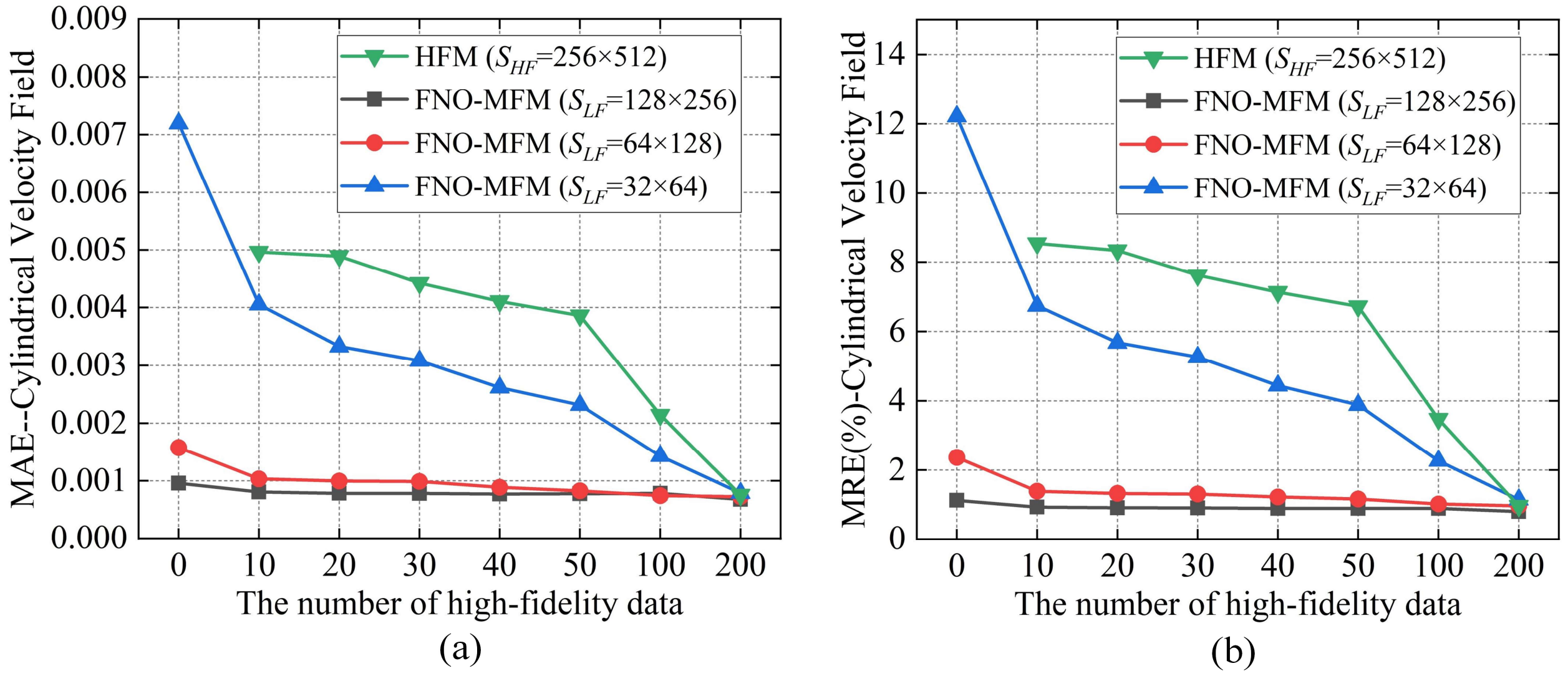

**Fig. 21.** Test error of single-cylinder wake versus the number of high-fidelity data: (a) MAE; (b) MRE.

**Table 4**

Error Reduction Error for laminar single-cylinder wake.

| $N^{high}$ | 10 | 20 | 30 | 40 | 50 | 100 | 200 |
|---|---|---|---|---|---|---|---|
| $ERR_{256\text{-}128}^{MAE}$ | 0.287 | 0.273 | 0.266 | 0.152 | 0.064 | 0.052 | 0.067 |
| $ERR_{128\text{-}64}^{MAE}$ | 2.894 | 2.333 | 2.108 | 1.947 | 1.804 | 0.925 | 0.085 |
| $ERR_{256\text{-}128}^{MRE}$ | 0.495 | 0.458 | 0.446 | 0.370 | 0.305 | 0.142 | 0.207 |
| $ERR_{128\text{-}64}^{MRE}$ | 3.856 | 3.290 | 3.027 | 2.636 | 2.344 | 1.233 | 0.190 |

c. Comparison between multi-fidelity model and high-fidelity model

In Fig. 21-Fig. 24, HFM with $S_{HF}$ = 256×512 is constructed for comparison. The FNO-MFMs have higher accuracy compared with the HFM under the same high-fidelity data. As shown in Fig. 22, the error ratios between FNO-MFM and HFM are all lower than the value of 1.0 except in a few cases. For all settings of $N_H$, the lowest MAE ratio of three FNO-MFMs are 0.16, 0.20, and 0.60, respectively; and the corresponding lowest MRE ratios are 0.11, 0.17, and 0.58, respectively. Moreover, the accuracy of FNO-MFMs with $S_{LF}$ = 128×256 and $S_{LF}$ = 64×128 when $N_H$ = 0 is higher than that of HFM when $N_H$ = 100. Obviously, for HFM, when $N_H$ = 10-100, although the vortex shedding is formed, the vortices are not in the correct wake position, and the predictive performance is staggeringly inefficient, as drawn in Fig. 23 and Fig. 24. In addition, HFM cannot realize the zero-data high-fidelity prediction. When $N_H$ = 0, no wake features are shown at all in the predictive velocity field, and the MAE of each point in the flow field here (the difference between the prediction and truth) is large.

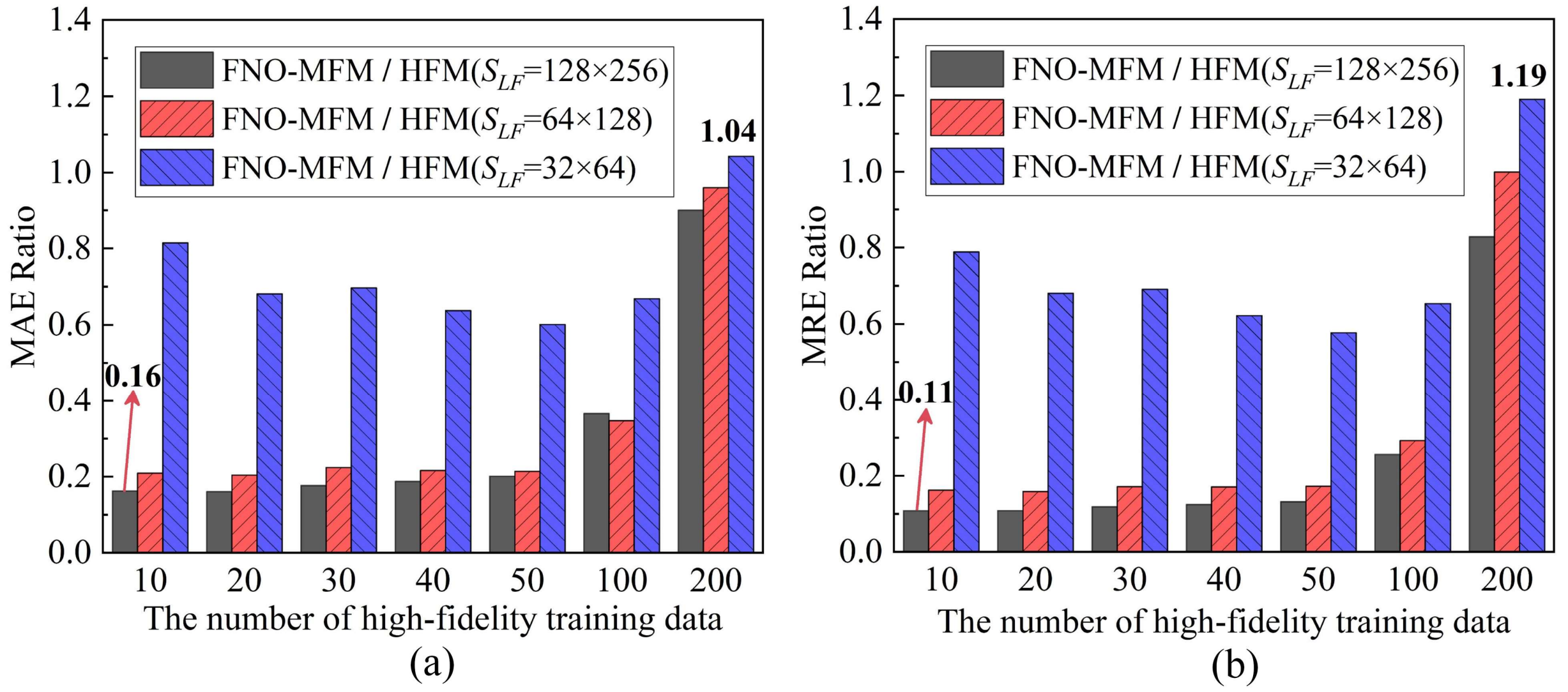

**Fig. 22.** Single-cylinder wake--ratio of test error between FNO-MFMs and HFM: (a) MAE; (b) MRE.

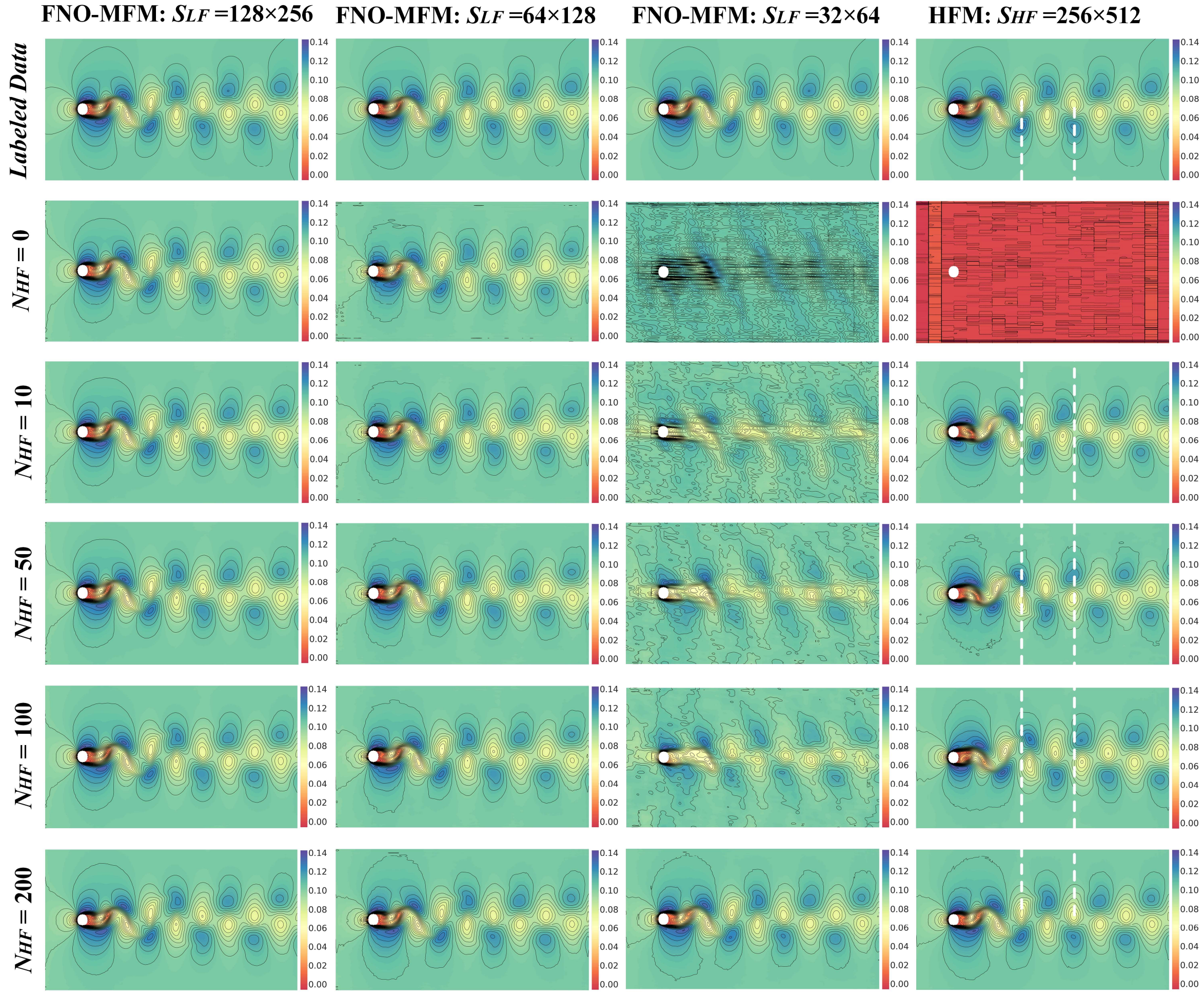

**Fig. 23.** Qualitative comparison of the performance in predicting the velocity field of laminar single-cylinder wake using three FNO-MFMs (FNO-MFM I: $S_{LF}$=128×256 and $S_{HF}$=256×512； FNO-MFM II: $S_{LF}$=64×128 and $S_{HF}$=256×512; FNO-MFM III: $S_{LF}$=32×64 and $S_{HF}$=256×512) and one HFM ($S_{HF}$=256×512).

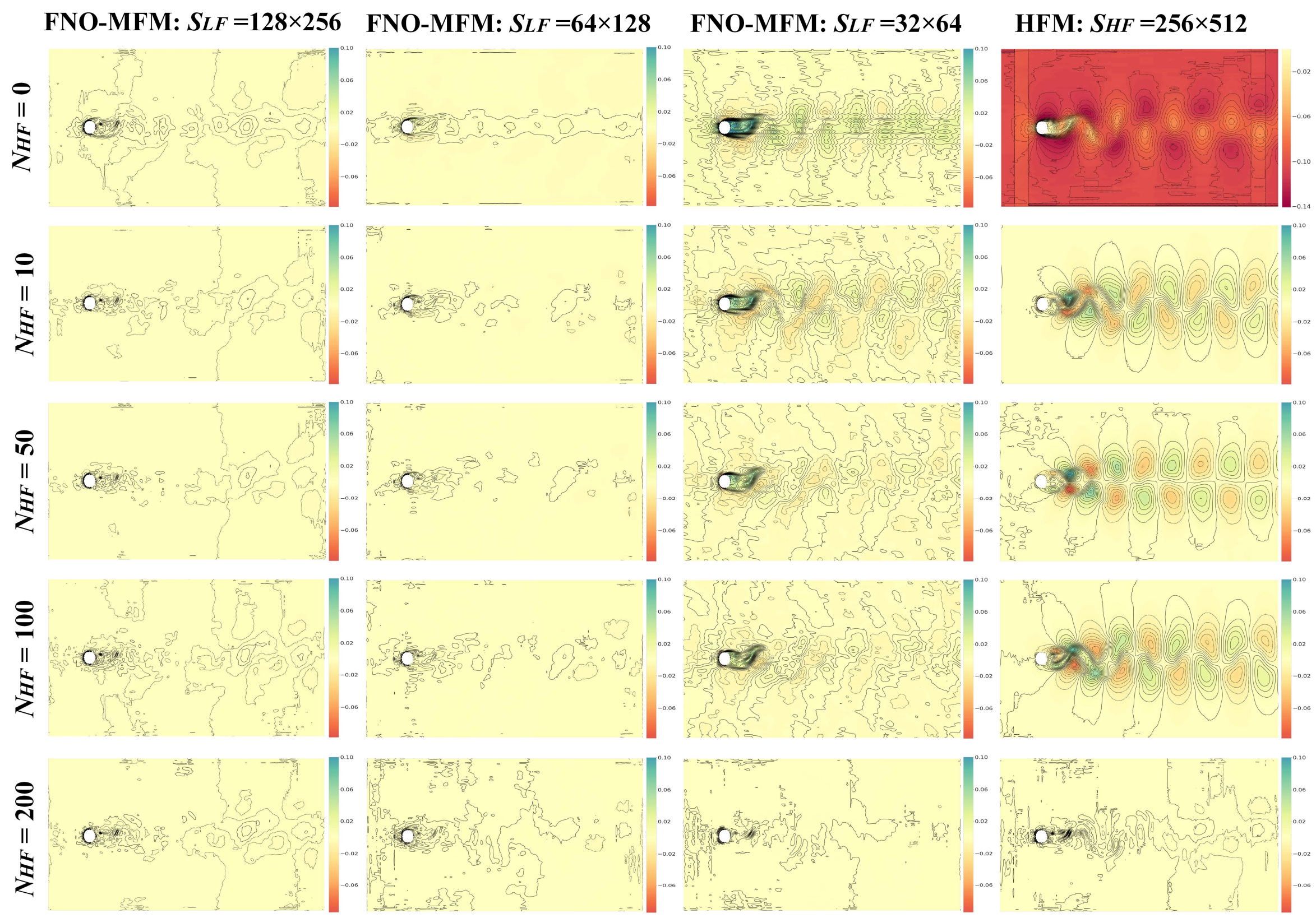

**Fig. 24.** Qualitative comparison of the MAE in predicting the velocity field of laminar single-cylinder wake using three FNO-MFMs (FNO-MFM I: $S_{LF}$=128×256 and $S_{HF}$=256×512； FNO-MFM II: $S_{LF}$=64×128 and $S_{HF}$=256×512; FNO-MFM III: $S_{LF}$=32×64 and $S_{HF}$=256×512) and one HFM ($S_{HF}$=256×512).

*4.5. Modeling timeliness of multi-fidelity learning method based on FNO*

Besides the three kinds of comparisons (i.e., a, b, c) in Section 4.4, another important index to evaluate the performance of FNO-MFM is the modeling timeliness. Modeling timeliness refers to the high precision performance of the FNO-MFM relatives to the HFM by training samples generated within the same time. In this section, the FNO-MFM of the temperature field is taken as an instance to explore and analyze the modeling timeliness.

Being different from the values identified in Section 4.3, the initial learning rates adopted by the pre-training and fine-tuning of FNO-MFMs here are selected as 0.001 and 0.005, respectively. The batch size and epochs are set to, respectively, 8 and 200. With 70 and 150 as the cut-off points, the learning rate of three epoch sections is 1, 0.1, and 0.01 times the initial value, respectively. As the total generation time of training samples is fixed at $T$ = 233s (containing 100 high-fidelity samples), the MAE on test data for different FNO-MFMs and HFM are compared in the form of points in Fig. 25 (a). It should be emphasized that each point represents the average MAE of ten different test processes under the same sample generation time. The variables in Fig. 25 (a) are the fidelity of FNO-MFMs and the different combinations of the number of low- and high-fidelity data. The fidelities of FNO-MFMs are $S_{LF}$ = 100×100, 50×50, and 25×25, respectively. The selected number of high-fidelity training data is composed of 9 values that are evenly distributed from 10 to 90. The corresponding amount of low-fidelity training data is dependent on the time left after stripping away the high-fidelity samples. Additionally, the HFM has the sample number of $N_H$ = 100 and the fidelity size of $S_{HF}$ = 200×200. The time required by the temperature field generator to obtain a single sample data of four different fidelities (in descending order) is 2.33s, 0.553s, 0.135s, and 0.055s, respectively.

It can be observed from Fig. 25 (a) that, as $T$ = 233s, three FNO-MFMs all have higher modeling accuracy than HFM, within a limited range of the number of high-fidelity samples. In ascending order of fidelity, this limited range is $N_H$=50-90, $N_H$=30-90, and $N_H$=10-70, respectively. More intuitively, this range has shifted to the direction with the lower number of high-fidelity data, and the number of high-fidelity samples corresponding to the lowest MAE for each FNO-MFM decreases under satisfying the timeliness. The latter phenomenon is reasonable on account of low-fidelity samples with higher fidelity containing more real physical field information. Moreover, the higher the fidelity of FNO-MFM, the lower the minimum MAE obtained. For instance, FNO-MFM with $S_{LF}$ = 100×100 has value of $MAE_{min}$ = 0.1435 (295LFS + 30HFS); FNO-MFM with $S_{LF}$ = 50×50 has value of $MAE_{min}$ = 0.15671 (518LFS + 70HFS); FNO-MFM with $S_{LF}$ = 25×25 has value of $MAE_{min}$ = 0.19765 (848LFS + 80HFS).

When the total time $T$ is doubled to 466s, as shown in Fig. 25 (b), meaning that the low- and high-fidelity samples double in quantitative terms. Among them, FNO-MFMs with $S_{LF}$ = 100×100 and 50×50 have a similar tendency as the case of $T$ = 233s, and have the minimum value of MAE of 0.0891 (422LFS + 100HFS) and 0.0855 (1036LFS + 140HFS), respectively. Nevertheless, FNO-MFM with $S_{LF}$ = 25×25 has a higher MAE than HFM.

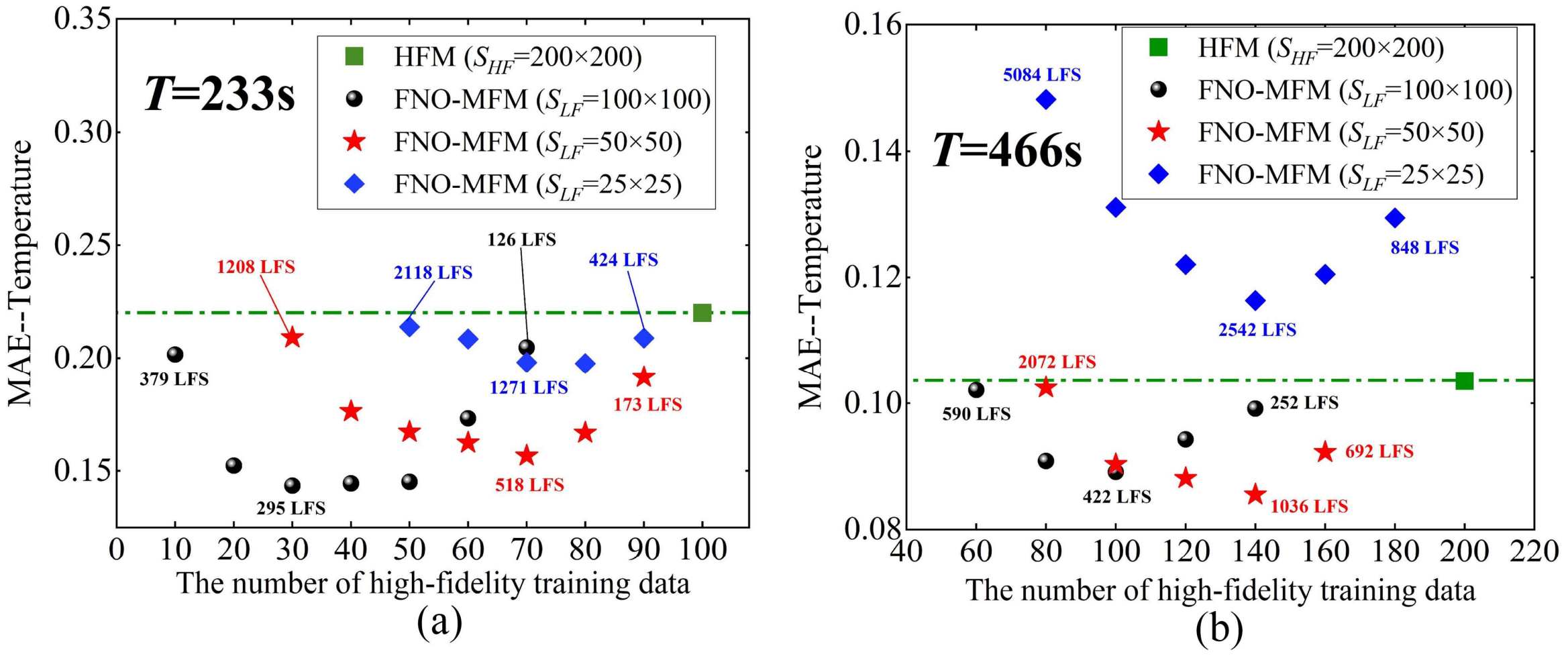

**Fig. 25.** The MAE on test data for different FNO-MFMs and HFM when the total sample generation time is fixed: (a) $T$ = 233s; (b) $T$ = 466s.

## 5. Conclusion

A data-adaptive multi-fidelity model based on FNO was proposed to replace the high-fidelity model by reducing the number of expensive high-fidelity data while maintaining the modeling accuracy. Combined with transfer learning, the establishment of a multi-fidelity model was realized by training large amounts of cheap low-fidelity data and a limited number of expensive high-fidelity data successively. To validate the effectiveness of the multi-fidelity model, three typical engineering examples were utilized to analyze the effect of the fidelity of low-fidelity data and the number of high-fidelity data on the performance. The advantages of the multi-fidelity model can also be represented by the corresponding high-fidelity model with the same amount of high-fidelity training data. Finally, the modeling timeliness of multi-fidelity modeling was explored. The following conclusions can be observed in this paper:

(1) For the single-fidelity model, with increasing Epoch, the validation loss of FNO-model has a faster-descending convergence rate and a smaller stable error than that of the CNN model, meaning that the selected basic neural network of the multi-fidelity model, i.e., FNO, is guaranteed for the modeling accuracy.

(2) For the three selected engineering examples, the FNO-based multi-fidelity model can reach up to satisfactory accuracy of 99%, and has higher predictive precision than that of the high-fidelity model under a fixed number of high-fidelity data.

(3) The higher the fidelity of low-fidelity data, the higher the accuracy of the FNO-based multi-fidelity model. For the temperature field, the minimum value of the MAE ratio and MRE ratio are 0.045 and 0.05, all of them belong to the FNO-MFM with $S_{LF}$ = 100×100; For the single-cylinder wake field with $S_{LF}$ = 128×256, these two minimum ratios are 0.16 and 0.11, respectively.
(4) Increasing the number of high-fidelity training data will improve the modeling accuracy.
(5) For modeling timeliness of temperature field models, when the total time of data generation is $T$ = 233s, three FNO-MFMs all have lower test MAE than that of HFM within a certain range of high-fidelity data volume. Although this phenomenon is weakened when $T$ increases from 233s to 466s, the FNO-MFMs with $S_{LF}$ = 100×100 and $S_{LF}$ = 50×50 are still better than the corresponding HFMs.
(6) Based on the conclusions that have been drawn, future work can focus on improving the accuracy of FNO-based multi-fidelity models by incorporating valid physical information into the models.

**Declaration of competing interest**

The authors declare that they have no known competing financial interests or personal relationships that could have appeared to influence the work reported in this paper.

**Data availability**

Data will be made available on request.

**Acknowledgements**

This work was supported by the National Natural Science Foundation of China (No. 62001502).